\documentclass[sn-mathphys-num]{sn-jnl}

\usepackage{graphicx}%
\usepackage{multirow}%
\usepackage{amsmath,amssymb,amsfonts}%
\usepackage{amsthm}%
\usepackage{mathrsfs}%
\usepackage[title]{appendix}%
\usepackage{xcolor}%
\usepackage{textcomp}%
\usepackage{manyfoot}%
\usepackage{booktabs}%
\usepackage{algorithm}%
\usepackage{algorithmicx}%
\usepackage{algpseudocode}%
\usepackage{listings}%
\usepackage{pifont}

\newcommand{\pdifft}[1]{\frac{\partial #1}{\partial t}}
\newcommand{\pdiffx}[1]{\frac{\partial #1}{\partial x}}

\theoremstyle{thmstyleone}%
\theoremstyle{thmstyletwo}%

\theoremstyle{thmstylethree}%

\begin{document}

\title[Article Title]{The impact of different degrees of leadership on collective navigation in follower-leader systems}


\author*[1]{\fnm{Sara} \sur{Bernardi}}\email{sara.bernardi@polito.it}

\author[2]{\fnm{Kevin J.} \sur{Painter}}\email{ kevin.painter@polito.it}

\affil[1]{\orgdiv{Department of Mathematical Sciences “G. L. Lagrange”}, \orgname{Politecnico di Torino}, \orgaddress{\street{Corso Duca degli Abruzzi 24}, \postcode{10129}, \state{Turin}, \country{Italy}}}

\affil[2]{\orgdiv{Interuniversity Department of Regional and Urban Studies and Planning}, \orgname{Politecnico di Torino}, \orgaddress{\street{Viale Pier Andrea Mattioli 39}, \postcode{10125}, \state{Turin}, \country{Italy}}}


\abstract{
In both animal and cell populations, the presence of leaders often underlies the success of collective migration processes, which we characterise by a group maintaining a cohesive configuration that consistently moves toward a target.
We extend a recent non-local hyperbolic model for follower-leader systems to investigate different degrees of leadership. Specifically, we consider three levels of leadership: \emph{indifferent} leaders, who do not alter their movement according to followers; \emph{observant} leaders, who attempt to remain connected with the followers, but do not allow followers to affect their desired alignment; and \emph{persuadable} leaders, who integrate their attempt to reach some target with the alignment of all neighbours, both followers and leaders. A combination of analysis and numerical simulations is used to investigate under which conditions each degree of leadership allows successful collective movement to a destination.
We find that the indifferent leaders' strategy can result in a cohesive and target-directed migration only for short times. Observant and persuadable leaders instead provide robust guidance, showing that the optimal leader behavior depends on the connection between the migrating individuals: if alignment is low, greater follower influence on leaders is beneficial for successful guidance; otherwise, it can be detrimental and may generate various unsuccessful swarming dynamics.}




\maketitle

\section{Introduction}\label{sec1}

The expansion of Austronesian and Micronesian populations across the south Pacific demanded, quite naturally, considerable prowess for maritime navigation. Many navigating cues would have been used, including celestial bodies, characteristic wave and wind patterns, and probably others since forgotten. Following the paths of certain seabirds, such as terns, to their home islands provided an important source of information to mariners, pointing them in the direction of landmasses \cite{lewis1994we}.

Follower-leader systems have attracted increasing levels of attention as a paradigm for certain collective movement processes, within both ecological \cite{berdahl2018collective,brent2015ecological,strandburg2018inferring} and cellular \cite{vishwakarma2018mechanical, cheung2013collective,vilchez2021decoding} populations. Simplistically, in a follower-leader driven process the leaders can be viewed as those members that establish and/or indicate a route that the followers can adopt. For example, leaders may be those with {\em a priori} information regarding the location of a destination, or through subtly different movement characteristics take up a more prominent position within the migrating group. However, while this partitioning into followers and leaders is conceptually straightforward, it remains a broad and indeterminate separation. In the opening example above, the leaders (seabirds) are highly distinct from the followers (mariners) and, most likely, indifferent to whether the followers remain within visual contact to successfully complete their navigation. In other instances, leaders may be highly vested in the migration success of the followers. African elephants matriarchs, for example, appear to provide leadership during group travel, offering guidance to water sources \cite{payne2003sources}. High-resolution GPS data reveal that dominant female meerkats exert greater influence than other individuals on both group direction and speed during foraging \cite{averly2022disentangling}.
The family structure inherent to such groups would exert a strong pressure to maintain group compactness, the leaders providing route guidance but also remaining attentive that followers stay on route. In other instances the distinction between followers and leaders may be even more subtle, for example within a population of relatively homogeneous individuals that differ in a phenotypic trait -- for example, speed, “courage”, or ability to detect navigating cues – which leads them to adopt a more prominent/position role in the overall guidance \cite{sasaki2018personality}. For example, a subtle leadership has been experimentally shown in shoals of sticklebacks, where the presence of partially informed individuals help them to find and access hidden food \cite{webster2017fish}. 

Numerous theoretical models have been developed to describe the collective movement of a population, ranging from agent-based models that track each individual’s movement path (e.g. \cite{aoki1982,reynolds1987flocks,vicsek1995novel,couzin2002collective,berdahl2018collective}) to integro-partial differential equation models for evolving population density distributions \cite{eftimie2018hyperbolic,painter2024biological}. Unifying these varied approaches is the assumption that the movement dynamics are partly driven by various interactions between neighbouring individuals \cite{carrillo2010particle}, typically some set of: individual to individual repulsion at short distances, e.g. to prevent collisions; individual to individual attraction at larger distances, e.g. to prevent group dispersal; and, individual to individual alignment, so that the group converges on a common orientation. Layered on top of this, external guidance cues may act to bias the overall movements towards some particular target \cite{codling2016balancing}. 

Recently, substantial research into collective behaviour has focused on the significance of population heterogeneity. Heterogeneity may occur at a heterospecific level -- grazing groups formed from zebras, gazelles, wildebeest \cite{saltz2023identifying} -- or within the same species. At a cellular level, migrating cell populations may be composed by cells of multiple types, or cells of the same type but structured into distinct phenotypes with distinct motility and proliferative behaviours \cite{giese1996dichotomy}. 
In the context of heterogeneous groups composed from a mixture followers and leaders, it is logical to assume that the various orientation influencing factors described above will vary between the two populations. For example, in the context of mariners following seabirds, while it may be natural to assume that the mariners attract and/or align according to the observed positions and trajectories of  birds, the birds are unlikely to respond similarly. In other follower-leader systems, where leaders “pay attention”, both followers and leaders may have some combination of attraction and alignment interactions according to all members of the population. It is also likely that leaders will differentiate themselves in other manners, such as through their capacity to detect and orient according to external guidance cues, adopting different speeds, or through controlling their level of conspicuousness. From a modeling perspective, the description of a broader range of follower-leader heterogeneity has been proposed using agent-based models. The balance between individual information and social influence among leaders in collective migration is studied in \cite{guttal2010social}, where the authors also explore migratory strategies in response to environmental threats. In \cite{couzin2011uninformed}, both modeling theory and experiments are used to highlight the role of followers in achieving democratic consensus, particularly in the presence of internal group conflicts and informational constraints.

In this paper we consider an abstract framework that allows us to explore the extent to which different “degrees” of leadership can impact on collective migration processes. In particular, we build on a non-local hyperbolic framework of follower-leader dynamics formulated in \cite{bernardi2021leadership}, which is in itself an extension of the model of Eftimie et al. \cite{eftimie2007modeling} to describe self-organisation and collective motility of a homogeneous population. Notably, this framework incorporates the various types of group-interaction described above, but through the continuous formulation remains somewhat amenable to formal analysis. In \cite{bernardi2021leadership} our focus was on the different mechanisms through which a leader population could impart directional information to a population of followers, but the leaders themselves were given a fixed set of interactions via which they interacted with the followers. Here we will relax that assumption and consider different types of leader. 

The remainder of this paper is organized as follows. In the next section, we will describe the various types of leaders we will consider, and lay out the mathematical framework. In Section \ref{sec3}, we introduce and characterize the dynamics of the \emph{indifferent} leader model. Sections \ref{sec4} and \ref{sec5} cover the formulation and resulting dynamics of the \emph{observant} and \emph{persuadable} leader models, respectively. We conclude with a discussion and suggestions for future investigation.

\section{Degrees of leader and model structure}\label{sec2}

\subsection{Degrees of leadership}

\begin{figure}
    \centering
   \includegraphics[width=0.8\textwidth]{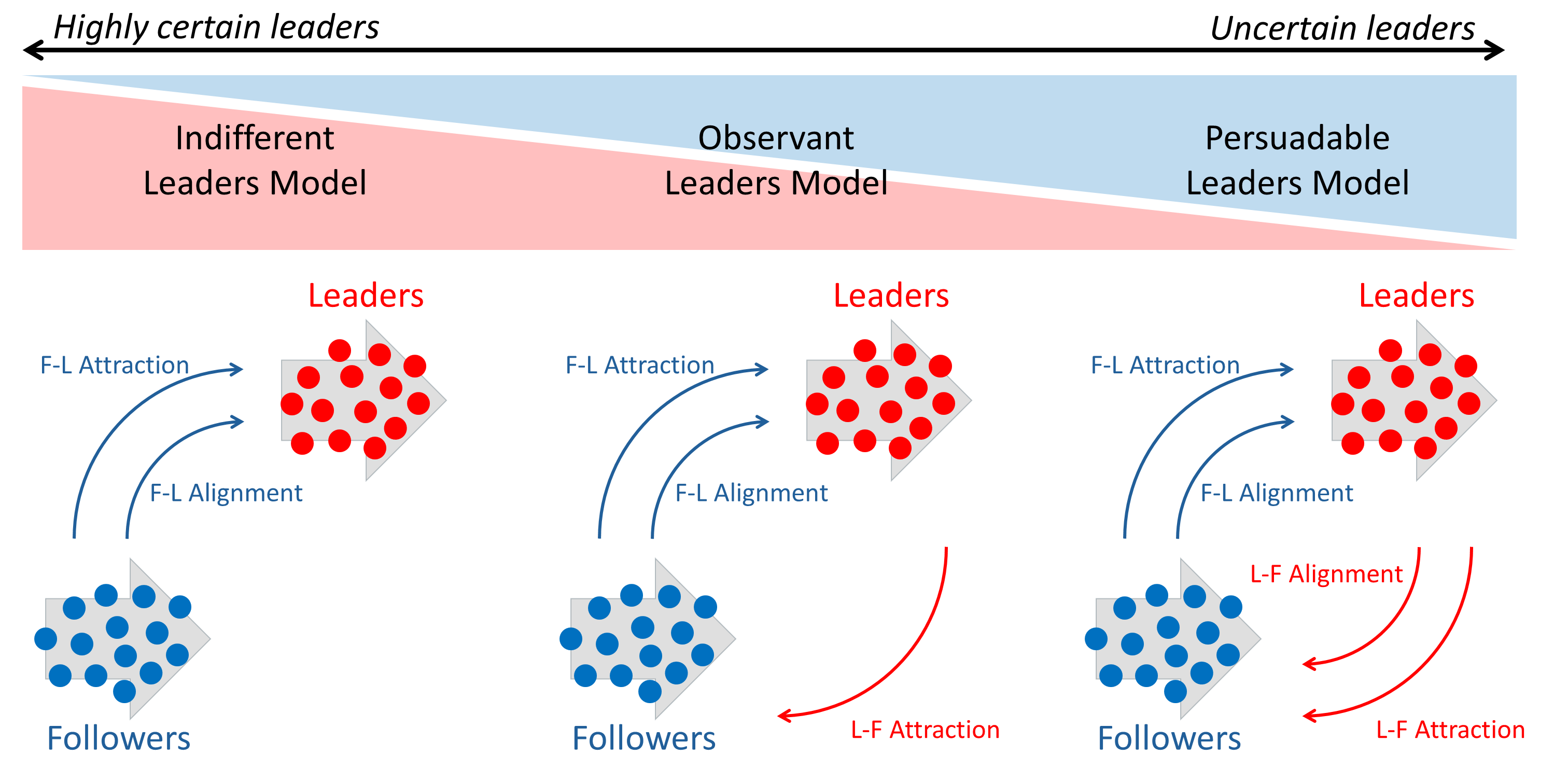}
    \caption{Modeling different degrees of leadership, ranging from highly certain leaders to uncertain leaders. \emph{Indifferent} leaders have no movement response to followers; \emph{observant} leaders respond to follower position through attraction; \emph{persuadable} leaders respond to both follower position and alignment.}
\label{hierarchy_of_models}
\end{figure}

As described in mathematical detail below, our intention is to formulate and analyse a model for follower-leader type collective guidance, where the movement dynamics of the two populations are determined by interactions between neighbouring individuals. Here we will restrict to a combination of
\begin{itemize}
\item {\em Attraction interactions}, in which individuals have a tendency to orient in the direction of higher population numbers. Attraction allows a swarm to form or maintain an aggregated state.
\item {\em Alignment interactions}, in which individuals have a tendency to align according to the dominating alignment. Alignment can allow polarisation, or consensus for a specific direction.
\end{itemize}
As noted earlier, repelling interactions can also be considered within collective movement models, although we exclude these here for model compactness. We assume that leaders are aware of the target direction (e.g. from {\em a priori} knowledge or detecting a cue) but followers are {\em na{\"i}ve}, and rely on the informed leaders. In previous work \cite{bernardi2021leadership} we assumed that leaders were somehow invested in the success of the followers, focussing on different mechanisms for imparting route information. As such, the precise manner by which followers influenced leader behaviour was kept fixed. Here we relax this, in particular by considering {\em three leadership degrees}, see Figure \ref{hierarchy_of_models}:
\begin{itemize}
\item {\em Indifferent} leaders. Followers have no influence on leader attractive or alignment responses. Thus, leaders are ``indifferent'' as to whether followers successfully follow.
\item {\em Observant} leaders. Followers influence leader attraction responses. We interpret this as a leader population that will alter  its movement in order to remain connected with followers, but do not allow followers to influence their own intended alignment. 
\item {\em Persuadable} leaders. Followers influence both leader attraction and alignment. This could reflect more ``subtle'' leaders that may have some information on the target, but can be swayed by the alignment choices of followers.
\end{itemize}

\subsection{Model framework} \label{sec:framework}

To implement these models we utilise a nonlocal hyperbolic PDE approach, first developed in Eftimie et al. in \cite{eftimie2007modeling} (see also \cite{eftimie2018hyperbolic}). Here we briefly summarise the model structure for the case of a single (homogeneous) population; in subsequent sections we outline its adaption to describe the various follower-leader systems. 

Consider a motile population, subdivided into its proportions moving right ($+$) or left ($-$), each with fixed speed $\gamma$. We set $p^+(x,t)$ and $p^-(x,t)$ to denote the population densities of the $+$ and $-$ populations, respectively, at position $x \in \Omega \subset \mathbb{R}$ and time $t \in [0, \infty)$. We let $p(x,t)=p^+(x,t)+p^-(x,t)$ denote the total population density. The governing equations are as follows:
\begin{eqnarray} 
\pdifft{p^+} + \gamma \pdiffx{p^+} & = & -\lambda^{p^+} p^+ +\lambda^{p^-} p^-\,, \nonumber\\
\pdifft{p^-} - \gamma \pdiffx{p^-} & = & +\lambda^{p^+} p^+ -\lambda^{p^-} p^-\,, \label{follower_model} \nonumber \\
p^{\pm}(x,0) & = & p_0^{\pm}(x).
\end{eqnarray}

Periodic boundary conditions are set in order to minimise the influence of the domain boundaries. 
The turning functions $\lambda^{p^+}$ and $\lambda^{p^-}$ describe the rate of switching direction, from right to left and left to right, respectively. These in turn depend (non-locally) on the distribution and orientation of perceived neighbours. As stated above, we consider two fundamental interaction types: (i) attraction, to encourage cohesion, and (ii) alignment, to encourage consensus of orientation. To implement this mathematically we first assume that the turning functions increase monotonically and smoothly depend on an internal control variable $y^{p^{\pm}}$. This, as we show below, integrates the population-level travel information perceived by each group member. Specifically, we set
\begin{equation}
\lambda^{p^\pm}=\lambda_1+\lambda_2 f(y^{p^\pm}), \quad \text{ where }
f(y)= 0.5+0.5 \tanh(y-y_0).
\end{equation}
The coefficients $\lambda_1$ and $\lambda_2$ denote the random and directed turning rate, respectively, and $y_0$ is chosen such that $f(0) \ll 1$. Next, variable $y^{p^\pm}$ is defined separately for the left and right moving populations as
\begin{equation}
y^{p^{\pm}}=\underbrace{Q_a^{p^{\pm}}}_{attraction}+\underbrace{Q_l^{p^{\pm}}}_{alignment},
\end{equation}
where larger (smaller) $y^{p^{\pm}}$ correspond to a higher (lower) tendency of switching direction. Finally, the social interaction terms $Q_a^{p^\pm}$ and $Q_l^{p^\pm}$ derive from the perceived positional and directional information of the neighbours. Specifically, attraction is based on the integrated total population density, such that turning is less likely if movement is in the direction of higher population densities:
\begin{equation}
    Q_a^{p^\pm} = -q_a \int_{0}^{\infty} K_a (s) \left(p(x \pm s)-  p(x \mp s) \right) ds,
\label{M0_attraction_follower}
\end{equation}
where $q_a$ and $K_a(s)$ denote the attractive strength and the interaction kernel, respectively. 

The alignment contribution will be different in various models, but is of the general form 
\begin{eqnarray}
Q_{l}^{p^\pm} = \pm q_{l} \int_0^{\infty} K_{l} (s) P(p^+,p^-) ds\,. \label{alignment_g}
\end{eqnarray}    
where $q_l$ and $K_l(s)$ represent the alignment strength and the interaction kernel, respectively, and $P(p^+,p^-)$ defines the portion of the population that affects the alignment. For a single population model, a suitable choice would be
\begin{equation}\label{F_alignment_M0}
	Q_{l}^{p^\pm} = \pm q_{l} \int_0^{\infty} K_{l} (s) [  p^{-} (x +  s) + p^{-} (x - s) - (p^{+} (x +  s) +  p^{+} (x - s))] ds\,.
\end{equation} 
Effectively, for individuals moving in the ($+$) direction, the above alignment contribution increases when the surrounding majority travels in the opposite ($-$) direction, and hence the rate of switching direction will be larger. 

Finally, we note that the expressions of $Q_l$ and $Q_a$ depend on interaction kernels
$K_l$ and $K_a$, which define how an interaction's strength depends on the separation distance. Following \cite{eftimie2007modeling}, we take these as translated Gaussians
\begin{equation} \label{kernels}
K_{i}(s)= \frac{1}{\sqrt{2 \pi m_{i}^2}} \exp \left(\frac{-(s-s_{i})^2}{2m_{i}^2} \right), \quad i=a, l  \quad s \in [0,\infty),
\end{equation}
where $s_a$ and $s_{l}$ are half the length of the attraction and alignment ranges, respectively. Note that the constants $m_{i}$, $i=a, l$, are chosen such that the support of more than $98\%$ of the mass of the kernels falls inside the interval $[0, \infty)$, i.e. $m_{i}=\frac{s_{i}}{8}$, $i=a, l$. This ensures the integral defined on $[0, \infty)$ in Eq. \eqref{alignment_g} to be a reasonably accurate approximation of that defined on the whole real line. A model built on the above lines was developed and analysed in \cite{eftimie2007modeling}. For further details, and some possible extensions, we refer to the book \cite{eftimie2018hyperbolic}.

\section{The \emph{indifferent leaders} model} \label{sec3}

\subsection{Formulation of the \emph{indifferent leaders} model}

An extension of models as described in (\ref{sec:framework}) to heterogeneous groups that contain both follower and leader types was considered in \cite{bernardi2021leadership}. Here we expand on that analysis to include different leader types. In the following, we will use $u^+$ and $u^-$ to denote the right and left moving follower subpopulations, and $v^+$ and $v^-$ to denote the right and left moving leader subpopulations. Consequently,  $u = u^+ + u^-$ and  $v = v^+ + v^-$ denote the total follower and leader densities, respectively. We further set the total population densities, $p^+ = u^+ + v^+$, $p^- = u^- + v^-$, and $p = u+v = u^+ + v^+ + u^- + v^-$.

The \emph{indifferent leader} model considers an extreme scenario in which leader behaviour is uninfluenced by the follower population. For example, these leaders may be entirely unrelated to followers and indifferent to their success (e.g. mariners following seabirds). To implement this in a simple manner, we introduce leaders as a compact population (assuming leaders maintain a group formation) that moves with constant speed ($\beta$) towards the right, which we subsequently refer to as the target direction. Specifically, 
\begin{eqnarray}\label{modelv}
	v^+(x,t)& = &M_{v}\exp(-0.5(x-x_0-\beta t)^2)\,, \nonumber \\
	v^-(x,t) & = &0 \,.
\end{eqnarray}
In the above, $M_v$ denotes the maximum leader density and $x_0$ indicates the initial position of the leader group. Follower dynamics, on the other hand,  are based on the system given by Eqs. \eqref{follower_model}:
\begin{eqnarray} \label{modelu}
\pdifft{u^+} + \gamma \pdiffx{u^+} & = & -\lambda^{u^+} u^+ +\lambda^{u^-} u^-\,, \nonumber\\
\pdifft{u^-} - \gamma \pdiffx{u^-} & = & +\lambda^{u^+} u^+ -\lambda^{u^-} u^- \,, \nonumber \\
u^{\pm}(x,0) & = & u_0^{\pm}(x)\,.
\end{eqnarray}
We follow the framework above by assuming the follower turning functions
\begin{equation}
\lambda^{u^\pm}=\lambda_1+\lambda_2 [0.5+0.5 \tanh(y^{u^\pm}-y_0)], \quad \text{ with } y^{u^\pm}=Q_a^{u^\pm} +Q_l^{u^\pm}.
\end{equation}
For the attractive contribution we assume as before an attraction according to the total population density, i.e.
\begin{equation}
    Q_a^{u^\pm} = -q_a \int_{0}^{\infty} K_a (s) \left( p(x \pm s) -  p(x \mp s) \right) ds \,. \label{attraction_follower}
\end{equation}
For the alignment, we adapt (\ref{F_alignment_M0}) to the form
\begin{equation}\label{F_alignment_M1}
Q_{l}^{u^\pm} = \pm q_{l} \int_0^{\infty} K_{l} (s) [  (u^{-} (x +  s) + u^{-} (x - s)) -
 (u^{+} (x +  s) +  u^{+} (x - s) + \alpha (v^+(x+s) +  v^+(x-s))) ] ds\,.
\end{equation} 
For $\alpha = 1$ the above is a straightforward generalisation of (\ref{F_alignment_M0}), but where followers now turn according to the prevailing orientation of the total (follower and leader) population. Choices of $\alpha \ne 1$, however, allow different levels of prioritisation: for example, a choice $\alpha > 1$ corresponds to a situation in which followers can discriminate between leaders and followers, and prioritise the orientation of the leaders in their alignment response.

\subsection{Dynamics of the \emph{Indifferent leaders} model}

Our  principal objective is to assess whether the presence of the (indifferent) leader population can provide sufficient information for a group of followers to travel towards the target (here, taken to be in the ``+'' direction). 
As such, we consider initial conditions in which the groups of followers and leaders initially overlap, but the follower population is essentially unbiased with respect to its orientation. Bearing in mind the imposed leader distributions of (\ref{modelv}), we therefore set
\begin{eqnarray}
	u^+(x,0)=\frac{M_{u}\exp(-0.5(x-x_0)^2)(1+r_u(x))}{2}, \\
	u^-(x,0)=\frac{M_{u}\exp(-0.5(x-x_0)^2)(1-r_u(x))}{2},
\end{eqnarray}
where $M_u$ denote the maximum follower density and $r_u(x)$ describes a random perturbation selected from a uniform distribution on $[-0.05, 0.05]$. Details of the numerical scheme are provided in \cite{bernardi2021leadership}.

A relatively basic set of requirements for ``success'' at the level of the follower population would be if the following two principal features are achieved: (i) that the followers achieve group consensus for the correct direction of travel, i.e. the majority of the population moves towards the target; (ii) that the followers maintain contact with the leading group, i.e. within a distance that allows transmission of information from leaders to followers. We note that (i) in the above list would be sufficient if the target direction was fixed and the leaders travelled in a straight line path to the target, but would be insufficient otherwise: maintaining contact with the leader group would therefore be necessary for robust guidance information.

\begin{figure}[htb!]
	\centering
	\includegraphics[width=\textwidth]{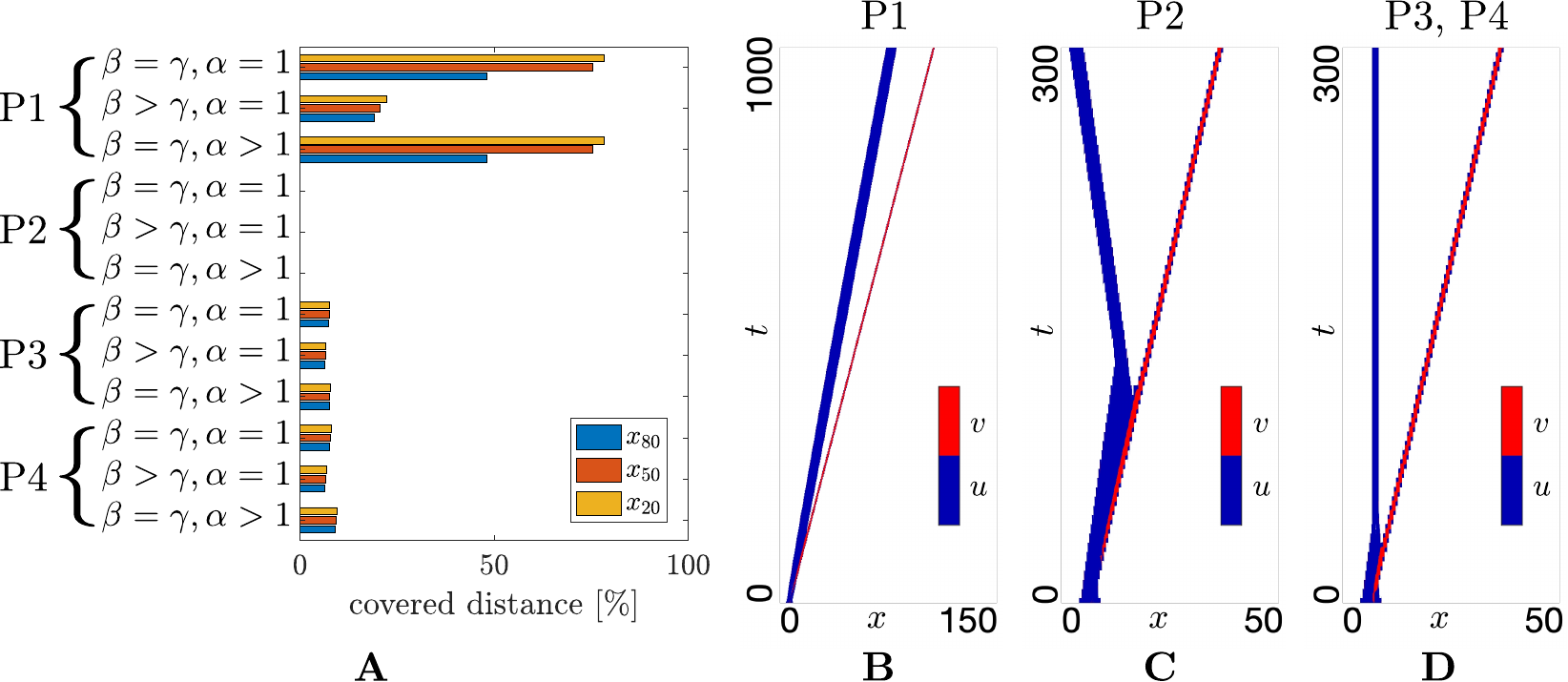}
	\caption{Dynamics of M1. (A) Second decile, median and eighth decile of the follower distribution for different parameter regimes: P1 weak attraction and strong alignment ($q_a=0.5, q_l=2$); P2 strong attraction and strong alignment ($q_a=2, q_l=2$); P3 strong attraction and weak alignment ($q_a = 2, q_l=0.1$); P4 weak attraction and weak alignment ($q_a = 0.5, q_l=0.1$). Results are shown for variations of key parameters describing speed differential between leaders and followers (i.e. $\beta > \gamma$) and leader influence on followers (i.e. $\alpha$). Specifically, we focus on: (i) $\beta=\gamma=0.1$, $\alpha=1$; (ii) $\beta=0.5$, $\gamma=0.1$, $\alpha=1$; (iii) $\gamma=\beta=0.1$, $\alpha=5$. (B-D) Space-time evolution of densities for $\beta=\gamma$, $\alpha=1$, under (B) P1, (C) P2, (D) P3, P4. Other parameter values are set as $\lambda_1=0.2$, $\lambda_2=0.9$, $M_u=M_v=12.61$, and $x_0=6.5$.}
	\label{M1_Figure}
\end{figure}

In Figure \ref{M1_Figure}A we describe the follower distribution across the spatial domain at the time instant when the leaders have migrated along a hundred space units. This is done by measuring the second decile, the median and the eigth decile of the follower distribution resulting from different parameter regimes. Specifically, we focus on four general parameter regimes: (P1) weak attraction and strong alignment, (P2) strong attraction and strong alignment, (P3) strong attraction and weak alignment and (P4) weak attraction and weak alignment. 

In the strong alignment regimes P1 and P2, the indifferent leader model can correctly align the followers on shorter timescales (see Figure \ref{M1_Figure}B and \ref{M1_Figure}C). However, over longer time scales followers lose contact with the leader group, generating different forms of failed collective migration. Under P1 we see an almost immediate splitting of the followers and leader groups, with the former traveling at a lower speed but still in the direction imparted through the brief initial contact with the leader population (see Figure \ref{M1_Figure}B). Clearly, this would be a problematic outcome if the target direction changed over time.
Under P2, the splitting of the two populations is followed by a change of movement direction of the follower individuals (see Figure \ref{M1_Figure}C). Here, as the leaders move beyond the range of perception of followers, the followers are subject to a strong attraction towards the backward of the swarm: this leads to a rapid change of heading which is then maintained due to the strong alignment interactions.
Under weak alignment regimes P3 and P4, the followers immediately lose contact with the leader group, leading to a splitting followed by followers coming to rest in the form of a stationary aggregate. Here the alignment within followers is insufficient to allow the group to polarise itself into a consensus for the group movement.
Note that for all the investigated parameter regimes we have a sufficient amount of follower to follower attraction to avoid group dispersion.
Figure \ref{M1_Figure}A also shows that results do not qualitatively change when a speed differential is introduced between leaders and followers (assuming leaders to move faster than followers to point the target direction - i.e., $\beta > \gamma$), as well as increasing the amount of leader influence on followers (i.e., $\alpha > 1$).
In summary, we find that indifferent leaders can guide follower migration towards the correct direction over short times, provided there is sufficient follower attraction and alignment strength. However, over longer times followers are generally unable to keep pace with the leaders, with the consequent emergence of different failure scenarios. Thus, the indifferent leader model does not appear to provide a robust guidance mechanism.

\section{\emph{Observant} and \emph{Persuadable leaders} models} \label{sec4}

To characterize the \emph{observant} and \emph{persuadable} leaders models, we introduce three means by which leaders are biased in the target direction. Building on our previous work \cite{bernardi2021leadership}, we assume that leaders have at least one of:
\begin{itemize}
    \item \emph{Orientation bias} (Obias), where leaders preferentially move in the target direction,
    \item \emph{Speed bias} (Sbias), where leaders move faster when target-directed,
    \item \emph{Conspicuousness bias} (Cbias),  where leaders moving towards the target are more conspicuous to followers than leaders moving away from the target.
\end{itemize}

\paragraph{Model 2, the Observant leaders model}

The \emph{observant leader} model is the first of two models in which leaders respond to follower distributions. Consequently, we extend \eqref{follower_model} to the following system of equations that couple follower ($u^+(x,t)$ and $u^-(x,t)$) and leader ($v^+(x,t)$ and $v^-(x,t)$) dynamics
\begin{eqnarray} \label{M2}
\pdifft{u^+} + \gamma \pdiffx{u^+} & = & -\lambda^{u^+} u^+ +\lambda^{u^-} u^-\, \nonumber\\
\pdifft{u^-} - \gamma \pdiffx{u^-} & = & +\lambda^{u^+} u^+ -\lambda^{u^-} u^- \, \nonumber \\
\pdifft{v^+} + \beta_+ \pdiffx{v^+} & = & -\lambda^{v^+} v^+ +\lambda^{v^-} v^- \, \nonumber \\
\pdifft{v^-} - \beta_- \pdiffx{v^-} & = & +\lambda^{v^+} v^+ -\lambda^{v^-} v^- \, \nonumber \\
u^{\pm}(x,0) & = & u_0^{\pm}(x) \nonumber \\
v^{\pm}(x,0) & = & v_0^{\pm}(x). 
\end{eqnarray}

We include a potentially different speed for leaders that are moving towards the target (i.e., $\beta_+ >\beta_-$), i.e. if Sbias is in operation. 


The turning functions are again given by
\begin{equation}
\lambda^{i^\pm}=\lambda_1+\lambda_2 [0.5+0.5 \tanh(y^{i^\pm}-y_0)], \quad \text{ with } y^{i^\pm}=Q_a^{i^\pm} +Q_l^{i^\pm}, \quad \text{ for } i \in \{u, v\}.
\end{equation}

We assume the attractive interaction to be active for both leader and follower individuals: all swarm members, regardless of their follower or leader status, prefer to maintain connection with neighbours. 
This assumption results in the following attraction terms:
\begin{equation}
    Q_a^{u^\pm} = Q_a^{v^\pm} = -q_a \int_{0}^{\infty} K_a (s) \left(u(x \pm s)+v(x \pm s)-  u(x \mp s)-  v(x \mp s) \right) ds\,.
\label{attraction_follower_leader}
\end{equation}

The alignment term is distinct according to  each subgroup, as leaders and followers rely on different information sources. Specifically, we assume that leaders bias their movement direction according to the target, while neglecting other group members: 
\begin{eqnarray}
Q_{l}^{v^\pm} = \mp q_{l} \int_0^{\infty} K_{l} (s) \eta  \, ds\,,
\end{eqnarray} 
where $\eta$ quantifies the strength of the \emph{orientation bias} (Obias). The above states that leader alignment is uninfluenced by other swarm members, depending only on a (spatially uniform and constant) cue when
the orientation bias is operating, i.e. when $\eta >0$.

Assumed to be completely uninformed regarding the destination, followers are instead taken to align according to the orientation of the  neighbours,
\begin{eqnarray}
Q_{l}^{u^\pm} = \pm q_{l} \int_0^{\infty} K_{l} (s) [  u^{-} (x +  s) + u^{-} (x - s) + \alpha^- (v^-(x+s) + v^-(x-s)) \nonumber \\- (u^{+} (x +  s) +  u^{+} (x - s)) - \alpha^+ (v^+(x+s) +  v^+(x-s)) ] ds\,.
\end{eqnarray} 

Note that the target direction can potentially be favoured through $\alpha^+ >\alpha^-$, i.e. the Cbias that makes leaders more conspicuous when moving towards the target.

\paragraph{Model 3, the Persuadable leaders model}

The \emph{Persuadable leader} model represents the most subtle level of leader behaviour in our modeling framework, and is obtained by assuming that leader orientation incorporates both a social source of information and the impact of an environmental cue, the latter described by the parameter $\eta$ (Obias), as above. 
The leader alignment term is thus the point of distinction from M2 (see Table \ref{fig1:modeling_sheme}) and is given by:
\begin{eqnarray}
Q_{l}^{v^\pm} = \pm q_{l} \int_0^{\infty} K_{l} (s) [ u^{-} (x +  s) + u^{-} (x - s) + \alpha^- (v^-(x+s) + v^-(x-s)) \nonumber \\- (u^{+} (x +  s) +  u^{+} (x - s)) - \alpha^+ (v^+(x+s) +  v^+(x-s)) - \eta ] ds\,.
\end{eqnarray}

We note that in M3 the same conspicuousness bias acts equally on leader and follower alignment terms. 
Table \ref{table1} summarizes the model parameters varied within the manuscript. We refer to Appendix \ref{fixedpar} for the models parameters that are kept at fixed values and set according to the previous works in \cite{eftimie2007modeling,bernardi2021leadership}.

\begin{table} 
\caption{Table of parameters varied within this study. 
The remaining model parameters are unchanged and set according to values from previous literature, see Appendix \ref{fixedpar}.
}
\label{table1}       
\begin{tabular}{lll}
\multicolumn{1}{c}{\textbf{Grouping}} & \multicolumn{1}{c}{\textbf{Parameter}} & \multicolumn{1}{c}{\textbf{Description}} \\
    \hline
Speed:   
&$\beta_+$          &speed of ($+$) moving leaders    \\   
                &$\beta_-$         &speed of ($-$) moving leaders  \\
              \hline
Attraction: &$q_a$     &attraction strength\\
              \hline
Alignment: &$q_l$     &alignment strength  \\
&$\alpha^+$            & alignment due to ($+$) oriented leaders  \\ 
       &$\alpha^-$            & alignment due to ($-$) oriented leaders\\ 
       &$\eta$         & environmental bias perceived by leaders\\ 
              \hline
Initial condition: &$x_0$ &Center position of aggregated initial configuration \\
\hline
Pop. size: &$A_u$ &mean follower density   \\
                     &$A_v$ &mean leader density  \\
                   &$M_u$ &maximum initial follower density\\
                     &$M_v$ &maximum initial leader density   \\
              \hline
Domain: &$L$ &domain length \\
\hline
Turning rates: &$\lambda_1$     &baseline turning rate\\
&$\lambda_2$     &maximum turning rate\\
\noalign{\smallskip}\hline
\end{tabular}
\end{table}

\begin{table}[h]
\caption{Modeling assumptions on the alignment influences for leaders and followers.  Followers do not receive information from the surrounding environment and adopt the predominant movement orientation among the neighbours. \emph{Observant} leaders align to the perceived environmental bias. \emph{Persuadable} leaders align to both the environmental cue and the rest of the group.}\label{tab1}%
\begin{tabular}{llll}
\toprule
Alignment sources & Followers  & Observant leaders & Persuadable leaders\\
\midrule
Direction of neighbours    & \ding{51}   & \ding{55}  & \ding{51}  \\
Environmental bias    & \ding{55}  & \ding{51}  & \ding{51}  \\
\botrule
\end{tabular}
\label{fig1:modeling_sheme}
\end{table}

\section{Dynamics of \emph{Observant} and \emph{Persuadable} leader models} \label{sec5}

As with the \emph{indifferent} leader model, we examine the conditions under which each of the \emph{observant} and \emph{persuadable} leader models ensures successful group migration, specifically by (i) aligning the heterogeneous population to the target direction and (ii) achieving coherent movement.
  
\subsection{Steady states and stability analysis of the non-spatial problem}

We first focus on the ability of the \emph{observant} and \emph{persuadable} leaders to generate consensus on group alignment towards the target, by performing a stability analysis of the non-spatial problem.
Specifically, we determine the spatially homogeneous steady-states $u^+(x,t)=u^*, u^-(x,t)=u^{**}, v^+(x,t)=v^*, v^-(x,t)=v^{**}$, with total density $A=A_u+A_v$, and examine their stability.

The steady-state set of equations for M2 reads as

\begin{eqnarray}
h_u^{\textup{M}2}(u^*, q_l, \lambda, A_u, A_v, \alpha^+, \alpha^-, y_0) &=& 0, \\
h_v^{\textup{M}2}(v^*, q_l, \lambda, A_v, \eta, y_0) &=& 0, \label{steady_state_eq_M2}
\end{eqnarray}
where 
\begin{eqnarray*}
h_u^{\textup{M}2}&=&-u^* \{ 1+\lambda \tanh \{2 q_l [ A_u-2u^* + \alpha^- (A_v - v^*)  - \alpha^+ v^*]-y_0 \} \} \nonumber  \\ &&+(A_u-u^*) \{1 + \lambda \tanh \{ -2 q_l [ A_u-2u^* + \alpha^- (A_v - v^*)  - \alpha^+ v^*] -y_0 \} \}, \\
h_v^{\textup{M}2}&=& - v^* [ 1+\lambda \tanh (-q_l \eta -y_0 ) ] + (A_v - v^*) [1+\lambda \tanh (q_l \eta -y_0 ) ],
\end{eqnarray*}

and $\lambda = \frac{0.5 \lambda_2}{0.5 \lambda_2 + \lambda_1}$.
 
From \eqref{steady_state_eq_M2}, we obtain a single steady-state $(v^*,v^{**})$ for the leader equation, where
\begin{equation}
v^*=\frac{A_v[1+\lambda \tanh( q_l \eta - y_0)]}{2+\lambda \tanh(-q_l  \eta -y_0) + \lambda \tanh(q_l  \eta - y_0)}, \quad v^{**}=A_v-v^*.   
\end{equation}

A similar set of equations is obtained for M3
\begin{eqnarray}
&h_u^{\textup{M}3}(u^*, q_l, \lambda, A_u, A_v, \alpha^+, \alpha^-, y_0) = h_u^{\textup{M}2}, \\
&h_v^{\textup{M}3}(v^*, q_l, \lambda, A_u, A_v, \alpha^+, \alpha^-, \eta, y_0) = 0,
\end{eqnarray}
where 
\begin{eqnarray*}
h_v^{M3}&=& - v^* \{ 1+\lambda \tanh \{ 2 q_l [  A_u-2u^* + \alpha^- (A_v - v^*)  - \alpha^+ v^*] -q_l \eta -y_0 \} \}  \\ &&+ (A_v - v^*) \{1+\lambda \tanh \{ -2q_l [A_u-2u^* +\alpha^- (A_v - v^*) - \alpha^+ v^*] +q_l \eta -y_0 \} \}.
\end{eqnarray*}


If there is no alignment, i.e. $q_l=0$, for both M2 and M3 we find the unaligned steady-state with both populations equally distributed in the ($\pm$) direction of movement, i.e. $(u^*, u^{**}, v^*, v^{**})=\left( \frac{A_u}{2}, \frac{A_u}{2}, \frac{A_v}{2}, \frac{A_v}{2} \right)$.
If $q_l \neq 0$, the steady-state expression is more intricate and to gain insight we focus on a few particular scenarios. 
Note that Sbias, i.e. $\beta_+ \neq \beta_-$, has no effect on the steady-state equations: we thus consider only how the other leader biases, i.e. Obias and Cbias, impact on steady states.

\begin{figure}[h!]
     {\includegraphics[width=\textwidth]{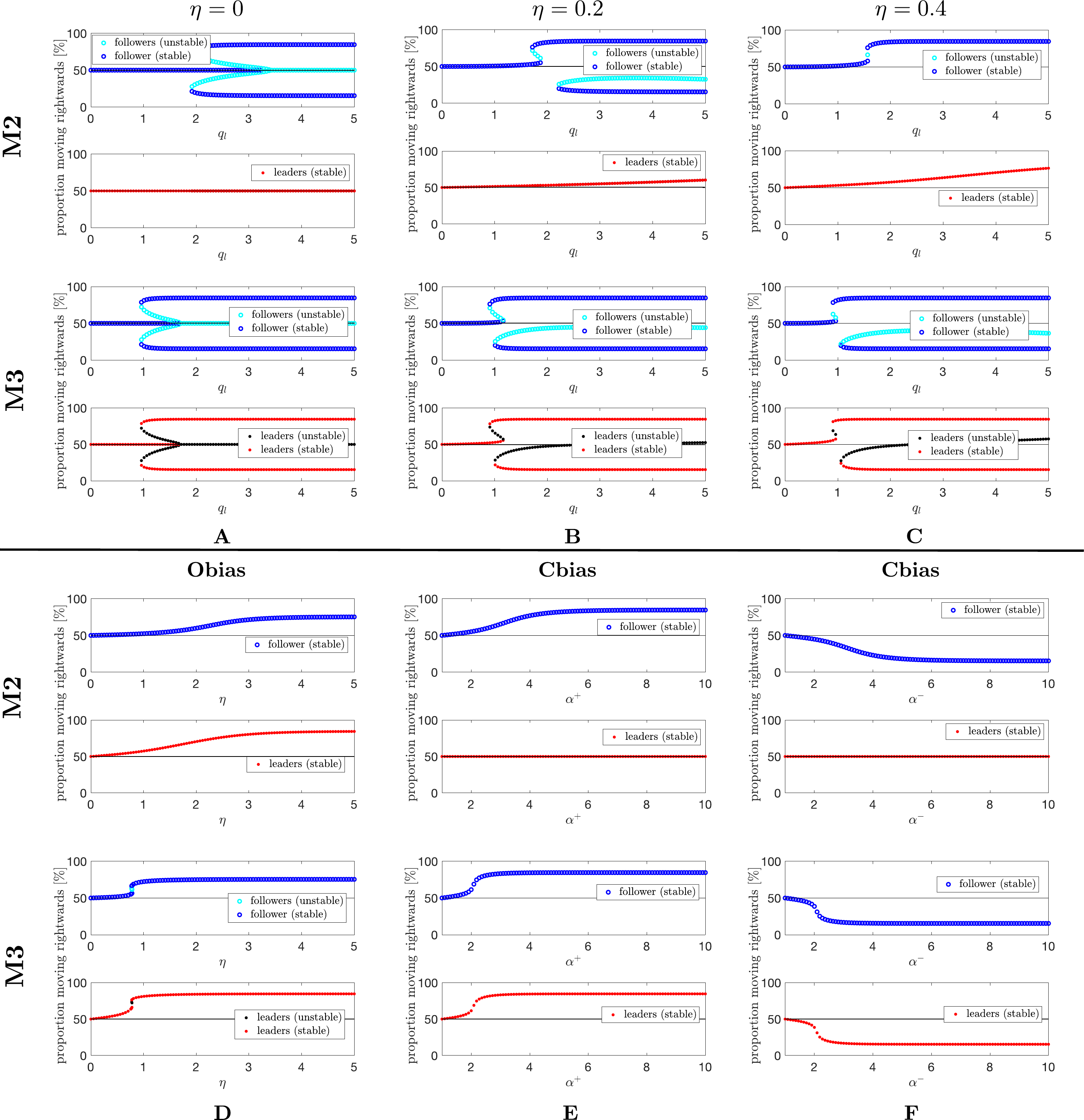}}
  \caption{M2 and M3, Proportion of right-moving populations at steady state(s). (A), (B), (C) Effect of $q_l$ on position and number of equilibrium points, for $\eta=0$, $\eta=0.2$,  $\eta=0.4$. (D) Effect of information level of leaders $\eta$ on position and number of equilibrium points, for $q_l=0.8$. (E), (F) Effect of the influence of right-moving (left-moving) leaders on followers $\alpha^+$ ($\alpha^-$) on position and number of equilibrium points, for $\alpha^-$ = 1 ($\alpha^+$ = 1) and $q_l=0.5$. Other parameter values fixed at $A_u = A_v = 1$, $\lambda_1 = 0.8$, and $\lambda_2 = 3.6$.}
 \label{ss_bif_diag}
 \end{figure}

The unbiased case is obtained when $\eta=0$, $\alpha^+=\alpha^-=1$. This reflects an absence of any leader source of movement information and in this case leaders are indistinguishable from followers. 
In M2, we find $(v^*, v^{**})= (\frac{A_v}{2}, \frac{A_v}{2})$ and the follower steady-state equation reduces to
\begin{equation}\label{ss_M2_unbiased}
h_u^{\textup{M}2}=-u^* \{ 1+\lambda \tanh \{2 q_l [ A_u-2u^*]-y_0 \} \} +(A_u-u^*) \{1 + \lambda \tanh \{ -2 q_l [ A_u-2u^* ] -y_0 \} \}. 
\end{equation}
In the context of a homogeneous population, the latter equation has been widely studied in \cite{eftimie2007modeling}. We restrict here to noting that the number of follower steady-states shifts between one, three and five, depending on the alignment strength $q_l$ (see Figure \ref{ss_bif_diag}A, top row). A sufficiently high alignment strength results in the stable follower steady state shifted to one in which a preferential direction of movement is selected (consensus). However, without any bias, consensus is symmetric with respect to both left ($-$) and right ($+$) directions.
In M3, under the same unbiased scenario, we find $h_u^{\textup{M}3}=h_v^{\textup{M}3}=h^{\textup{M}3}$: leader and follower steady-states coincide (see Figure \ref{ss_bif_diag}A, bottom row) and become the solution of the following
\begin{eqnarray}\label{ss_M3_unbiased}
h^{\textup{M}3}&=&-u^* \{ 1+\lambda \tanh \{2 q_l [ A_u-2u^* +  A_v - 2v^*)]-y_0 \} \} \nonumber \\
&&+(A_u-u^*) \{1 + \lambda \tanh \{ -2 q_l [ A_u-2u^* +  A_v  - 2v^*] -y_0 \} \}. 
\end{eqnarray}
The symmetry of the system obtained in the unbiased scenario is compromised when $\eta > 0$ (see Figure \ref{ss_bif_diag}B and Figure \ref{ss_bif_diag}C). The upper branches (corresponding to the target direction) are more likely to be selected, while the lower ones are shifted to the right and completely disappear for larger values of $\eta$. 
This is obtained for both M2 and M3; however, while consensus can be achieved in M3 at lower values of $q_l$ compared to M2, M3 requires much higher values of $\eta$ to make the lower branch disappear. This reflects that the orientation of persuadable leaders for M3 also stems from the orientation of followers, which, in turn, contribute to the determination of leader direction of movement. This process helps in reinforcing the target direction when $q_l$ is limited. However, higher values of $q_l$ amplify the information coming from the neighbours and larger values of $\eta$ are required in M3 to overcome the social influence.
For M2, the single leader steady-state is also shifted towards the (+) direction when $\eta>0$.

We also investigate the effect of increasing Obias strength when Cbias is inactive ($\eta \neq 0, \alpha^+=\alpha^-=1$). For both M2 and M3, if the alignment is not enough to provide a preferential alignment of the population, a small amount of environmental bias for leaders, $\eta \ne 0$, provides consensus. This is the situation depicted in Figure \ref{ss_bif_diag}D, where leaders and followers are equally distributed in left and right moving orientation when $\eta=0$.
Under the extreme Obias scenario, i.e. $\eta \to \infty$, both \emph{observant} and \emph{persuadable} leaders in M2 and M3 favour the (+) direction, i.e. $(v^*, v^{**})=(\frac{A_v(1+\lambda)}{2}, \frac{A_v(1-\lambda)}{2})$, and the follower steady-state equation reduces to
\begin{eqnarray} \label{ss_extremeObias}
h_u^{M2}=h_u^{M3}&=&-u^* \left\{ 1+\lambda \tanh \left\{2 q_l \left[ A_u-2 u^*- A_v \lambda \right]-y_0 \right\} \right\}  \nonumber  \\ &&+(A_u-u^*) \left\{1 + \lambda \tanh \left\{ -2 q_l \left[ A_u-2 u^*- A_v \lambda \right] -y_0 \right\} \right\}.
\end{eqnarray}
The symmetric structure of Eqs. \eqref{ss_M2_unbiased} and \eqref{ss_M3_unbiased} is therefore lost due to the quantity $A_v \lambda$, which acts to favour a shift of the steady-states towards the biased direction. 
Note also that in M3 a lower value of $\eta$ with respect to M2 is sufficient to reach consensus. Again, this is due to the positive feedback between leaders and followers which favours the transfer of the target direction of movement, provided a limited alignment strength $q_l$ ($q_l = 0.8$ in Figure \ref{ss_bif_diag}D).

We then turn to examine the effect of Cbias when Obias is inactive ($\alpha^+ / \alpha^-, \alpha^- / \alpha^+ > 0$, $\eta=0$). 
If the influence of right-moving (left-moving) leaders is sufficiently large the follower population will largely adopt a rightward (leftward) orientation movement, even in absence of the environmental bias, see Figure \ref{ss_bif_diag}E and \ref{ss_bif_diag}F.
Indeed, for both M2 and M3, the follower steady-states are shifted towards the (+) direction.  The same shift is observed for leader steady-state in M3 while in M2 leaders settle into the unaligned one (see Figure \ref{ss_bif_diag}E and \ref{ss_bif_diag}F). Consistently, we remark that $h_v^{M2}=0$ is not affected by Cbias.
Even if the leaders are equally distributed in orientation, a sufficiently high conspicuousness of their target-directed proportion provides consensus. This situation biologically occurs in honeybee swarming behaviour, where leaders are observed to fly back and forth within the swarming cloud with noticeable fast streaks when target directed.
For $\alpha^+ / \alpha^- \to \infty$, we find

$$(u^*, u^{**},v^*, v^{**})=\left(\frac{A_u(1+\lambda)}{2}, \frac{A_u(1-\lambda)}{2},\frac{A_v}{2}, \frac{A_v}{2}\right) \text{ for M2 and }$$

$$(u^*, u^{**},v^*, v^{**})=\left(\frac{A_u(1+\lambda)}{2}, \frac{A_u(1-\lambda)}{2},\frac{A_v(1+\lambda)}{2}, \frac{A_v(1-\lambda)}{2}\right) \text{ for M3. }$$ 






To summarise, bifurcation diagrams in Figure \ref{ss_bif_diag}  corroborate these findings and allows us to highlight the differences between the leader behaviour described in M2 and M3 in terms of the key parameters underlying the information transfer from leaders to followers, i.e. the alignment strength $q_l$, the environmental bias $\eta$, and the influence of target-directed leaders on followers quantified by the ratio $\frac{\alpha^+}{\alpha^-}$. Colors refers to the stability properties of the steady states (details on stability analysis of the time-only problem are included in the Appendix \ref{appendixA}).
The greater follower influence on the \emph{persuadable} leader behaviour described in M3 generally facilitate the coordination process: a lower value of $q_l$, 
$\eta$, and $\alpha^+ / \alpha^-$ with respect to M2 is indeed sufficient to reach consensus.
On the other hand, this positive feedback may lead the group to choose the wrong direction when alignment is too high (see Figure \ref{ss_bif_diag}A, \ref{ss_bif_diag}B and \ref{ss_bif_diag}C (bottom row)). 
These results thus suggest that M3 expands the parameter region where consensus is obtained, provided an optimal amount of alignment is set.
Moreover, we have shown the effect of the crucial parameters on the configuration of the state states. Specifically, $q_l$ generates different solutions structured in orientation while the bias parameters, i.e. $\eta$ and $\alpha^\pm$, shift them to one preferential direction of orientation.

\begin{figure}[h!]
   {\includegraphics[width=\textwidth]{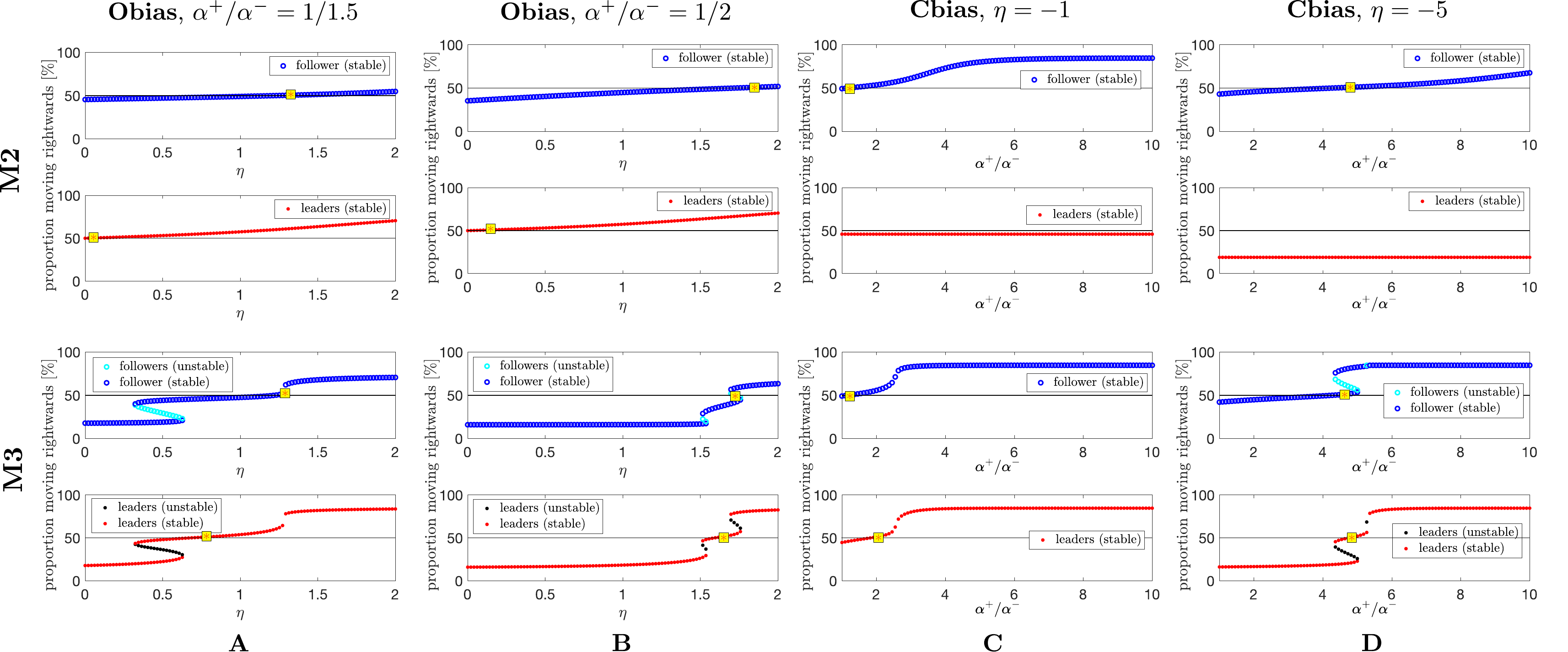}}
 \caption{M2 and M3, Proportion of right-moving populations at steady state(s). 
 (A-B) Effect of $\eta>0$ on position and number of equilibrium points: (A) $\frac{\alpha^+}{\alpha^-} = \frac{1}{1.5}$, (B) $\frac{\alpha^+}{\alpha^-} = 0.5$, with $q_l=0.8$.
 (C-D) Effect of $\frac{\alpha^+}{\alpha^-} > 0$ on position and number of equilibrium points: (C) for $\eta=-1$, (D) $\eta=-5$, with $q_l=0.5$. Other parameter values fixed at $A_u = A_v = 1$, $\lambda_1 = 0.8$, and $\lambda_2 = 3.6$. Note that the yellow square with a red star indicates the position where the steady state is symmetric with respect to the proportion of rightward v leftward oriented individuals.}
 \label{ss_bif_diag_confl_info}
 \end{figure}

\paragraph{Effect of conflicting information on leader guidance}
Now we discuss the effect of conflicting information (i.e. $\eta < 0$ and $\frac{\alpha^+}{\alpha^-} > 1$, and viceversa) on the position and number of steady-states, see Figure \ref{ss_bif_diag_confl_info}.
When leaders are (even slightly) more conspicuous when moving to the left (i.e. $\frac{\alpha^+}{\alpha^-}<1$), the effect of an environmental bias to the right, i.e. $\eta>0$, is consequently reduced in both M2 and M3, and larger values of $\eta$ are needed to have the (+) direction favoured, see Figure \ref{ss_bif_diag_confl_info}A and \ref{ss_bif_diag_confl_info}B. 
The opposite situation is depicted in Figure \ref{ss_bif_diag_confl_info}C and \ref{ss_bif_diag_confl_info}D. In presence of a negative environmental bias of leaders, i.e. $\eta<0$, if the conspicuousness of (+)-oriented leaders is sufficiently large, a consensus may be reached anyway. This occurs also in M2 where leader population settle into a predominantly (-)-oriented steady-state, regardless of the Cbias strength.
Generally, the stronger the Obias (Cbias) to the left, the larger the influence of right-moving leaders (right environmental leader bias) needs to be in order to reach a consensus where right orientation is favoured.
Moreover, for a fixed rightwards Obias (Cbias) strength, the $\alpha^+ / \alpha^-$ ($\eta$) threshold value required to generate consensus is higher in M2 w.r.t. M3, confirming the effect of reinforcement of information for \emph{persuadable} leaders (M3).
Interestingly, hysteresis loops are observed in M3 by varying $\eta$ (Figure \ref{ss_bif_diag_confl_info}A and \ref{ss_bif_diag_confl_info}B) and $\alpha^+$ (Figure \ref{ss_bif_diag_confl_info}D), suggesting that decreasing the bias parameter from a state of consensus may not lead to the group realigning to its previous state, e.g. potentially due to learned behavior.


\subsection{Numerical simulations}

\begin{figure}[H]
    \centering
\includegraphics[width=\textwidth]{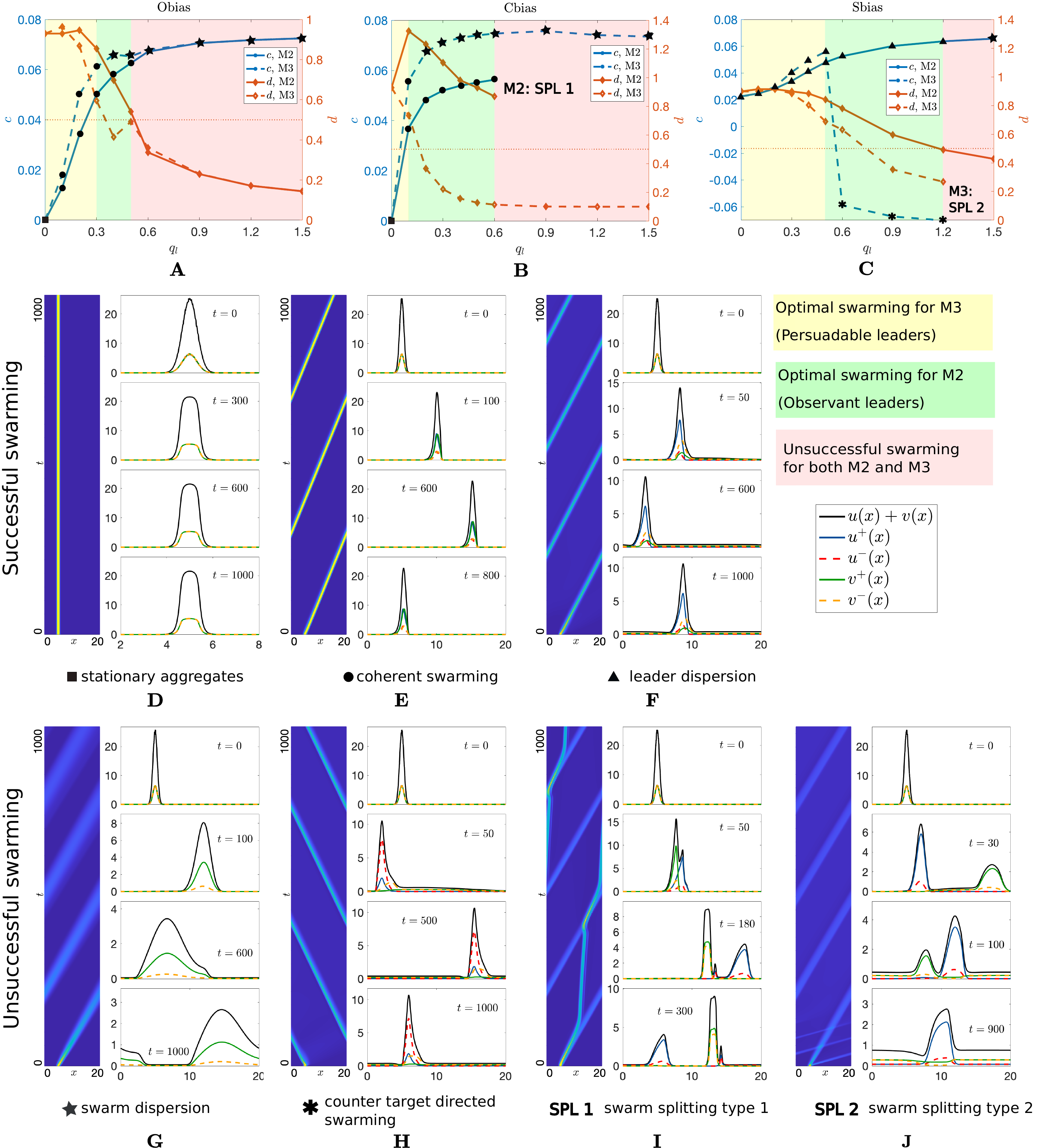}
    \caption{M2 and M3, Effect of Obias, Sbias and Cbias leader strategies on the swarm dynamics as $q_l$ increases under low attraction regimes and clustered initial configuration. (A-C): swarm dynamics is evaluated in terms of the speed (in blue) and the cohesion index (in orange) of the follower population. We highlight in yellow the $q_l$ values for which optimal swarming is obtained for M3, in green those for which optimal swarming is obtained for M2. Red regions denote unsuccessful swarming for both M2 and M3. Numerical simulations are obtained for (A) $\eta=10$, $\alpha^\pm=1$, $\beta_\pm=0.1$, (B) $\beta_+/\beta_-=0.5/0.1$, $\eta=1$, $\alpha^\pm=1$, (C) $\alpha^+/\alpha^-=5/1$, $\beta_\pm=0.1$, $\eta=1$. (D-F): examples of successful swarming patterns displayed by M2 and M3 (see correspondence with panels in the first row). (G-J): examples of unsuccessful swarming patterns displayed by M2 and M3 (see correspondence with panels in the first row). Other parameter values are set as $q_a=0.5$, $\lambda_1=0.2$, $\lambda_2=0.9$, $M_u=12.61$, $M_v=12.61$, and $x_0=5$.}
    \label{results_M2_M3_qa_05}
\end{figure}

The steady state and stability analyses of the non-spatial problem have revealed insight into the alignment process provided by the the \emph{observant} and \emph{persuadable} leader models. We now incorporate the spatial variable into the problem to investigate their ability to generate a spatially coherent group migration. In this regard, a numerical approach is necessary to explore the complex dynamics in space.
In Figure \ref{results_M2_M3_qa_05} we test the efficiency of different guidance strategies (Obias, Cbias and Sbias) when adopted by observant and persuadable leaders, respectively described by M2 and M3. 
Given that the point of distinction between M2 and M3 lies in the leaders' alignment terms, we examine how the resulting collective behavior changes with variations in the alignment strength. 
The other parameter values selected for the numerical realizations are chosen such that the linear stability analysis of the uniform solution predicts Turing pattern formation (the emerging aggregates are predicted to attain a spatially grouped configuration) in the unbiased case, i.e. when no leader bias is in operation, see Appendix \ref{appendixLSA}.  
In this respect, an attraction strength $q_a$ above a certain threshold turns out to play a key role in aggregating the population, which is of particular interest in the context of collective movement. Specifically, sufficiently strong attraction allows the populations to form and maintain a grouped form.

When introducing the effect of the three different leader strategies, we will refer to successful swarming dynamics if the follower group (i) moves in the direction of the target from their origin and (ii) maintains a compact group (i.e., follower individuals are not lost). Accordingly, we describe the resulting dynamics in terms of two relevant quantities. First, we look at the mean speed of the follower aggregation over the whole observation time. Secondly, we track a ``cohesion index'' measured as $d=\frac{d_0}{d_{\textup{end}}}$, where $d_0$ and $d_{\textup{end}}$ respectively denote the spatial extension $d=x_{80}-x_{20}$ of the follower aggregation measured at the initial time instant and at the end of the observation time, being $x_{20}$ and $x_{80}$ the second and the eighth deciles of the follower distribution. In this respect, $d \approx 1$ will reflect that a good coherence is maintained during the follower migration; conversely, $d \ll 1$ reflects follower dispersion.

The results show a clear trade-off: as the alignment strength increases, the speed of the groups increases while the cohesion index decreases for both M2 and M3. Furthermore, for lower values of $q_l$ all guidance strategies under M3 generates higher speed for the migrating swarm compared to M2 (see Figure \ref{results_M2_M3_qa_05}, first row, yellow regions). For higher values of $q_l$, M2 instead offers a better swarming, as for M3 we observe a drop in the cohesion index under the action of Obias and Cbias, while Sbias generates a counter-directed swarming (see Figure \ref{results_M2_M3_qa_05}, first row, green regions).
Even higher values of $q_l$ lead to unsuccessful swarming behaviour for both M2 and M3, including swarm dispersion, counter target directed swarming and two types of swarm splitting (with the leaders leaving the followers behind, and vice versa). 

M2 and M3 thus generate a variety of successful and unsuccessful swarming dynamics and we discuss some of the swarming scenarios below. 
Due to being incorporated within the alignment term, both Obias and Cbias can lead to stationary aggregates when there is no alignment, i.e. $q_l=0$ suppresses Obias and Cbias action and the swarm self-organizes into a compact cluster through attraction, see Figure \ref{results_M2_M3_qa_05}A, Figure \ref{results_M2_M3_qa_05}B and Figure \ref{results_M2_M3_qa_05}D. If $q_l=0$, Sbias can however provide a coherent (though slow) follower migration towards the target, while leaders gradually disperse, see Figure \ref{results_M2_M3_qa_05}C and Figure \ref{results_M2_M3_qa_05}F. Indeed, leaders with a greater speed towards the target are able to broadcast traveling information to the followers through attraction but are not able to self maintain a compact cluster, due to being both attracted both to slower follower and other faster leader individuals.
When $q_l > 0$, the action of Obias and Cbias in M2 and M3 
turns out to be the best guidance process with both leader and follower populations compactly moving to the destination, provided the alignment strength is not excessively strong, see Figure \ref{results_M2_M3_qa_05}A, Figure \ref{results_M2_M3_qa_05}B (green and yellow regions), and Figure \ref{results_M2_M3_qa_05}E. When $q_l$ passes above a certain threshold, under M3 both Obias and Cbias can lead to dispersion of both leader and follower populations, see Figure \ref{results_M2_M3_qa_05}A, Figure \ref{results_M2_M3_qa_05}B (red regions) and Figure \ref{results_M2_M3_qa_05}G. The same result is observed under M2 for an  Obias strategy, while Cbias generates swarm splitting of type I. In this case, leaders evolve to a stationary configuration and the followers split away from them in the direction of the target, being favoured from the initial propulsive thrust received from leaders, see Figure \ref{results_M2_M3_qa_05}B (red region) and Figure \ref{results_M2_M3_qa_05}I.
Similarly, if alignment is too high then Sbias provides a failure swarming scenario with swarm dispersion under M2, see Figure \ref{results_M2_M3_qa_05}C (red region), and swarm splitting of type II under M3, with faster leaders leaving the followers behind, see Figure \ref{results_M2_M3_qa_05}C (red region) and Figure \ref{results_M2_M3_qa_05}J.

Under a higher attraction regime, we find a successful swarming dynamics for all the investigated $q_l$ values (see Figure \ref{results_M2_M3_qa_2}, Appendix \ref{appendixExtraFigures}). In particular, increasing the attraction strength expands the parameter regions that represent best swarming for M3 (yellow regions) and M2 (green regions), respectively. This highlights the positive effect of attraction between group members on the guidance success.

To summarise, our numerical study suggests that when the alignment connection between all group members is limited, greater follower influence on leaders is beneficial for successful guidance. On the contrary, if the alignment connection is strong enough, a greater follower influence on leaders can be detrimental and may result in unsuccessful swarming.

\section{Discussion}\label{discuss}

Follower-leader heterogeneity often underlies the success of collective migration processes and occurs at different spatial and temporal scales, from cellular to animal systems \cite{qin2021roles,jolles2017consistent,pettit2015speed}. Such systems are characterized by the presence of knowledgeable and/or experienced leaders, who guide the movement direction for the naive followers. However, this clear-cut division overlooks a broader range of heterogeneity that can characterize the composition of the group.
In this respect, we have investigated different degrees of leadership, ranging from highly certain leaders to uncertain leaders, extending the recent non-local
hyperbolic model for follower-leader systems  proposed in \cite{bernardi2021leadership}. Specifically,
we have considered three degrees of leadership: indifferent leaders, who have no response to other group members; observant leaders, who show an attraction response to the neighbours, and persuadable leaders, who have both an attraction and an alignment response towards neighbours. Focusing on the ability of the resulting leader behaviour to (i) broadcast the target direction to the follower and (ii) keep the migrating group cohesive without dispersion, we have found that indifferent leaders do not provide successful guidance for long times, leading to group splitting. Thus, a kind of social connection from leaders to followers seems to be required for a successful guidance.
The importance of balancing goal-oriented and social-oriented behaviours for leaders is further supported by the experimental observations in \cite{ioannou2015potential}, where trained-to-target fish that exhibit high goal orientation -- characterized by faster and straighter paths to target -- and low social behaviour are likely to quickly reach the preferred target but tend to leave untrained fish behind, resulting in a failure to transmit their preference to others. 

Both observant and persuadable leaders instead provide robust guidance, showing that the best leader strategy depends on the alignment connection between the migrating individuals: a greater follower influence on leaders generally helps the guidance process when the alignment connection between individuals is limited; otherwise, it may generate an excess amount of directional information and lead to unsuccessful swarming dynamics. This result emerges both analytically, from the steady-state of the non-spatial problem and its stability properties, and numerically from simulations under moderate parameter variations. Moreover, it appears robust to variations in the three modeled mechanisms through which leaders can be biased towards the target direction: orientation (Obias), increased speed (Sbias), and conspicuousness (Cbias). 

Although the model is not designed to describe any specific system, but rather to provide general insights, examples from animal collective movement substantiate our findings. Observant leadership, for instance, is evident in migrating schools of fish and swarming honeybees. Specifically, target-trained fish demonstrate both individual knowledge of the target (to reach the food reward) and a strong tendency to maintain group cohesion \cite{miller2013both}. Informed scout bees perform high-speed flights through the migrating swarm while maintaining contact with the majority of uninformed bees \cite{schultz2008mechanism}. Similarly, persuadable leader behavior, which pools partial estimates of the homeward route with social information, is key to the effective migration of pigeon flocks \cite{herbert2015collective,pettit2015speed}.

The study offers deep insights into different leader behaviors, yet some simplifying assumptions have been required by the complexity of the modeling framework.
Since one of the key factors affecting the individual ability to gain directional information is the surroundings (e.g. weather condition,  physical landscape), the environmental conditions under which the migration occurs may be crucial in determining which is the best leader behaviour to be adopted -- specifically, the extent to which the leader's behavior gain benefit in being influenced by the rest of the group. In this regard, the influence of the environment in shaping individual and/or behavioral heterogeneity has been observed in both animal and cellular systems \cite{jolles2020role,mclennan2012multiscale}. An illustrative example is neural crest cell migration during embryogenesis. Studies here have shown the importance of the chemoattractant VEGF, not only in providing the direction of invasion but also in determining follower or leader status: cells exposed to higher VEGF gradients become leader cells, while those in lower gradients become followers \cite{mclennan2015vegf}. In this respect, it would be interesting to explicitly incorporate environmental effects into the model, relaxing the assumption of fixed individual behavior and instead making it depend on an environmental variable (e.g., quantifying the effect of a specific chemical factor, weather condition, or predation risk). Developing the model in this direction would require additional switching terms to account for transitions between different leader behaviors, and potentially between follower and leader modes as well. In this respect, transient leadership has already been investigated using kinetic models, with the leader-follower status described as either a discrete trait \cite{albi2024kinetic} or a continuous trait \cite{cristiani2024kinetic}.

\begin{figure}[htb]
    \centering
\includegraphics[width=\textwidth]{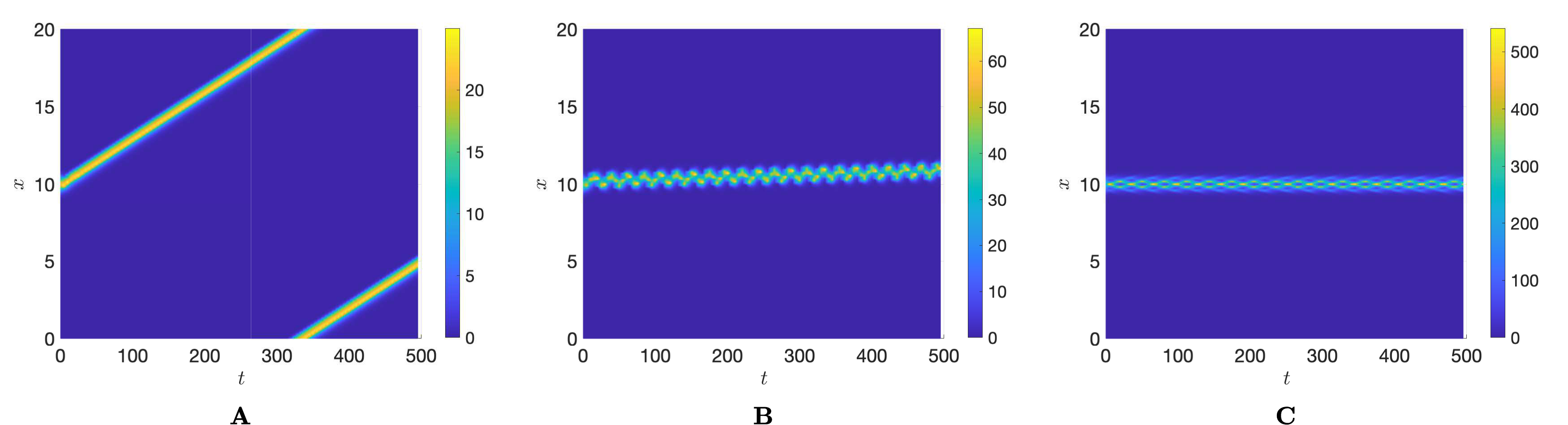}
    \caption{Effect of increasing the total follower population on the swarming dynamics described by M3. (A) Moving pattern obtained for $M_u=10$, (B) pulsating-moving pattern obtained for $M_u=30$, (C) pulsating-stationary pattern obtained for $M_u=200$, leaving unchanged $M_v=10$. Other parameters are set as $\gamma=\beta_{\pm}=0.1, q_a=2, q_l=0.5, \eta=0.5$, and $\alpha^\pm=1$.}
    \label{pulsating}
\end{figure}

We also note that this study has focused on fixed total densities of leaders and followers, which have been set to be equal. In this regard, preliminary numerical results suggest that increasing the follower population size leads to the emergence of pulsating-type patterns where the target-directed group migration provided by the leaders (specifically, Obias) can be lost (see Figure \ref{pulsating}). A more in-depth study of these dynamics would be valuable (e.g., to explore how large the follower population can be while still being successfully led by a given leader population), though it falls outside the scope of this work.
Another relevant aspect is that we have kept the interaction ranges of the follower and leader populations fixed and equal. Recent work in \cite{painter2024variations} has shown the impact of variations in non-local interaction ranges on the emergence of another form of group heterogeneity, namely chase-and-run dynamics. In chase-and-run systems, individuals of the first population move towards those of the second population, which, in turn, move away. Similar to follower-leader systems, these dynamics occur in populations of non-locally interacting cells or animals. In \cite{painter2024variations}, the authors show that chase-and-run dynamics are strongly influenced by non-local interaction terms and occur when the sensing range of the chaser exceeds that of the runner. Building on this, it would be interesting to investigate whether varying the sensing ranges of leader and follower populations could lead to similar results in the modeling framework used here. 

The model could also be adapted to study the co-presence of competing leaders behaviours pointing to different directions in a decision-making perspective.
In this respect, the dynamics emerging from the presence of two subsets of informed individuals, each having its own directional preference, is studied using a discrete approach in \cite{couzin2005effective}.

Furthermore, a promising direction would be the integration of data—such as measurements of speed, movement direction, and tracking data from migrating cell or animal populations—into the proposed model, in order to tailor it and gain insights into specific systems. Despite the rapidly increasing availability of such data thanks to modern technologies, incorporating them into hyperbolic models remains an open research challenge, as also discussed in \cite{eftimie2018hyperbolic}.

In conclusion, these models offer a solid step towards a broader investigation on the leader-follower heterogeneity underlying a variety of biological and ecological processes, offering insights on the individual mechanisms that may drive biological dynamics observed at the population level.



\bmhead{Acknowledgements}

KJP acknowledges ‘Miur-Dipartimento di
Eccellenza’ funding to the Dipartimento di Scienze, Progetto e Politiche del Territorio (DIST). SB acknowledge the financial support of GNFM-INdAM through ‘INdAM – GNFM Project’, CUP E53C22001930001. KJP and SB are members of INdAM-GNFM.





\begin{appendices}

\section{Stability analysis of the non-spatial problem}
\label{appendixA}

The time-only problem of models M2 and M3 presented in Eq. \eqref{M2} reduces to
\begin{eqnarray} \label{time_only_model}
\pdifft{u^+}  & = & -\lambda^{u^+} u^+ +\lambda^{u^-} u^-\,= f(u^+, u^-, v^+, v^-), \nonumber\\
\pdifft{u^-} & = & +\lambda^{u^+} u^+ -\lambda^{u^-} u^- \,= g(u^+, u^-, v^+, v^-), \nonumber \\
\pdifft{v^+} & = & -\lambda^{v^+} v^+ +\lambda^{v^-} v^- \,= p(u^+, u^-, v^+, v^-), \nonumber \\
\pdifft{v^-}  & = & +\lambda^{v^+} v^+ -\lambda^{v^-} v^- \, = q(u^+, u^-, v^+, v^-).
\end{eqnarray}

To assess the stability properties of the steady states, we evaluate the Jacobian matrix for the system \eqref{time_only_model} at a generic point:

\begin{equation}
J=\begin{pmatrix}
\frac{\partial f}{\partial u^+} & \frac{\partial f}{\partial u^-} & \frac{\partial f}{\partial v^+} & \frac{\partial f}{\partial v^-} \\
\frac{\partial g}{\partial u^+} & \frac{\partial g}{\partial u^-} & \frac{\partial g}{\partial v^+} & \frac{\partial g}{\partial v^-} \\
\frac{\partial p}{\partial u^+} & \frac{\partial p}{\partial u^-} & \frac{\partial p}{\partial v^+} & \frac{\partial p}{\partial v^-} \\
\frac{\partial q}{\partial u^+} & \frac{\partial q}{\partial u^-} & \frac{\partial q}{\partial v^+} & \frac{\partial q}{\partial v^-} 
\end{pmatrix}.
\end{equation}

Straightforward algebra leads to the following expressions for $\frac{\partial i}{\partial j^\pm}$, for $i \in \{f, g, p, q\}, j \in \{u, v\}$:

\begin{eqnarray*}
\frac{\partial f}{\partial u^+}&=& -\frac{\partial g}{\partial u^+}  =-\frac{\partial \lambda^{u^+}}{\partial u^+} u^+ - \lambda^{u^+} + \frac{\partial \lambda^{u^-}}{\partial u^+} u^-, \\
\frac{\partial f}{\partial u^-}&=&-\frac{\partial g}{\partial u^-}  =-\frac{\partial \lambda^{u^+}}{\partial u^-} u^+  + \frac{\partial \lambda^{u^-}}{\partial u^-} u^- + \lambda^{u^-}, \\
\frac{\partial f}{\partial v^+}&=& -\frac{\partial g}{\partial v^+}  =-\frac{\partial \lambda^{u^+}}{\partial v^+} u^+  + \frac{\partial \lambda^{u^-}}{\partial v^+} u^-,  \\
\frac{\partial f}{\partial v^-}&=& -\frac{\partial g}{\partial v^-}  = -\frac{\partial \lambda^{u^+}}{\partial v^-} u^+  + \frac{\partial \lambda^{u^-}}{\partial v^-} u^-, \\
\frac{\partial p}{\partial u^+}&=& -\frac{\partial q}{\partial u^+}  =-\frac{\partial \lambda^{v^+}}{\partial u^+} v^+  + \frac{\partial \lambda^{v^-}}{\partial u^+} v^-, \\
\frac{\partial p}{\partial u^-}&=&-\frac{\partial q}{\partial u^-}  =-\frac{\partial \lambda^{v^+}}{\partial u^-} v^+  + \frac{\partial \lambda^{v^-}}{\partial u^-} v^-,  \\
\frac{\partial p}{\partial v^+}&=& -\frac{\partial q}{\partial v^+}  =-\frac{\partial \lambda^{v^+}}{\partial v^+} v^+  -\lambda^{v^+}+ \frac{\partial \lambda^{v^-}}{\partial v^+} v^-,  \\
\frac{\partial p}{\partial v^-}&=& -\frac{\partial q}{\partial v^-}  = -\frac{\partial \lambda^{v^+}}{\partial v^-} v^+  + \frac{\partial \lambda^{v^-}}{\partial v^-} v^- +\lambda^{v^-}. \\
\end{eqnarray*}

In the above, for the \emph{Observant} leader model (M2) we have
\begin{eqnarray*}
\frac{\partial \lambda^{u^+}}{\partial u^\pm}&=& \mp \lambda_2 \, q_l \, [1-\tanh^2(y^{u^+}_{M2}-y_0)], \\
\frac{\partial \lambda^{u^+}}{\partial v^\pm}&=& \mp \lambda_2 \, q_l \, \alpha^{\pm} [1-\tanh^2(y^{u^+}_{M2}-y_0)], \\
\frac{\partial \lambda^{u^-}}{\partial u^\pm}&=& \pm \lambda_2 \, q_l \, [1-\tanh^2(y^{u^-}_{M2}-y_0)], \\
\frac{\partial \lambda^{u^-}}{\partial v^\pm}&=& \pm \lambda_2 \, q_l \, \alpha^{\pm} [1-\tanh^2(y^{u^-}_{M2}-y_0)], \\
\frac{\partial \lambda^{v^+}}{\partial u^\pm}&=& 
\frac{\partial \lambda^{v^+}}{\partial v^\pm}= 0, \\
\frac{\partial \lambda^{v^-}}{\partial u^\pm}&=&  
\frac{\partial \lambda^{v^-}}{\partial v^\pm}= 0, \\
\end{eqnarray*}
where the perceived signal function is given as:
\begin{eqnarray*}
y^{u^{\pm}}_{M2}&=&Q_l^{u^\pm}= \pm 2 \, q_l (  u^- + \alpha^- v^- - u^+ -\alpha^+ v^+). 
\end{eqnarray*}

For the \emph{Influenced} leader model (M3) we have
\begin{eqnarray*}
\frac{\partial \lambda^{u^+}}{\partial u^\pm}&=& \mp \lambda_2 \, q_l \, [1-\tanh^2(y^{u^+}_{M3}-y_0)], \\
\frac{\partial \lambda^{u^+}}{\partial v^\pm}&=& \mp \lambda_2 \, q_l \, \alpha^{\pm} [1-\tanh^2(y^{u^+}_{M3}-y_0)], \\
\frac{\partial \lambda^{u^-}}{\partial u^\pm}&=& \pm \lambda_2 \, q_l \, [1-\tanh^2(y^{u^-}_{M3}-y_0)], \\
\frac{\partial \lambda^{u^-}}{\partial v^\pm}&=& \pm \lambda_2 \, q_l \, \alpha^{\pm} [1-\tanh^2(y^{u^-}_{M3}-y_0)], \\
\frac{\partial \lambda^{v^+}}{\partial u^\pm}&=& \mp \lambda_2 \, q_l \, [1-\tanh^2(y^{v^+}_{M3}-y_0)], \\
\frac{\partial \lambda^{v^+}}{\partial v^\pm}&=& \mp \lambda_2 \, q_l \, \alpha^{\pm} [1-\tanh^2(y^{v^+}_{M3}-y_0)], \\
\frac{\partial \lambda^{v^-}}{\partial u^\pm}&=& \pm \lambda_2 \, q_l \, [1-\tanh^2(y^{v^-}_{M3}-y_0)], \\
\frac{\partial \lambda^{v^-}}{\partial v^\pm}&=& \pm \lambda_2 \, q_l \, \alpha^{\pm} [1-\tanh^2(y^{v^-}_{M3}-y_0)], \\
\end{eqnarray*}
where the perceived signal functions are given as:
\begin{eqnarray*}
y^{u^{\pm}}_{M3}&=&Q_l^{u^\pm}= \pm 2 \, q_l (  u^- + \alpha^- v^- - u^+ -\alpha^+ v^+), \\
y^{v^{\pm}}_{M3}&=&Q_l^{v^\pm}= \pm 2 \, q_l (  u^- + \alpha^- v^- - u^+ -\alpha^+ v^+-\frac{1}{2} \eta).
\end{eqnarray*}

If at least one of the eigenvalues has a positive real part, solutions near the steady state $S=(u^*,u^{**},v^*,v^{**})$ diverge and the steady state is unstable. Otherwise, solutions near $S$ converge to $S$ and the steady state is asymptotically stable.

\section{Linear stability analysis of the spatial problem}\label{appendixLSA}

To further assess the stability properties of the system and in particular the possibility of pattern formation, we perform a standard linear stability analysis on the \emph{Observant} and \emph{Persuadable} leader-only submodels, i.e. where we set $u^+=u^-=0$ in the corresponding systems of equations \cite{murray2003mathematical}. 
We note that extending the kernel $K_a$ to an odd function on the whole real line, Eqs. \ref{attraction_follower_leader} can be written as
\begin{eqnarray}
Q_a^{u^\pm} = -q_a \int_{-\infty}^{\infty} K_a(s) u(x \pm s) ds \\
Q_a^{v^\pm} = -q_a \int_{-\infty}^{\infty} K_a(s) v(x \pm s) ds.
\end{eqnarray}

We then set $v^+(x,t)=v^*+ v_p(x,t)$ and $v^-(x,t)=v^{**}+ v_m(x,t)$, where $v_p(x,t)$ and $v_m(x,t)$ each denote small perturbations. We substitute in the corresponding models, neglect nonlinear terms in $v_p$ and $v_m$ and look for solutions $v_{p,m} \propto e^{\sigma t + ikx}$. 
Some algebra leads to the dispersion relation of the form
\begin{equation} \label{disp_rel}
    \sigma^+(k)= \frac{C(k)+\sqrt{C^2(k)-D(k)}}{2},
\end{equation}
where the expression for $C(k)$ and $D(k)$ slightly change for each of the models considered.

Specifically, for the \emph{Observant} leader-only submodel (M2), we find
\begin{eqnarray*}
C(k)&=&(\beta_- - \beta_+)ik -2\lambda_1-\lambda_2-0.5\lambda_2 [\tanh(q_l \eta-y_0) +\tanh(-q_l\eta-y_0)], \\
D(k)&=& 4 \beta_+ \beta_- k^2 +4ik\lambda_1(\beta_+-\beta_-)+ 2ik \lambda_2 \{
\beta_+(1+\tanh(q_l\eta -y_0)) \\ &&
-\beta_-(1+\tanh(-q_l\eta -y_0)) 
+v^*(1-\tanh^2(-q_l \, \eta-y_0))[\beta_{+} q_a \hat{K}_a^+ + \beta_{-} q_a \hat{K}_a^+] \\ &&
+ v^{**}(1-\tanh^2(q_l \, \eta -y_0))[- \beta_+ q_a \hat{K}_a^- - \beta_- q_a \hat{K}_a^-] \}.
\end{eqnarray*}

Whereas, for the \emph{Influenced} leader-only submodel (M3), 
\begin{eqnarray*}
C(k)&=&(\beta_- -\beta_+)ik -2\lambda_1-\lambda_2-0.5\lambda_2 [\tanh(-2 q_l(\alpha^- v^{**}-\alpha^+ v^*-0.5 \eta)-y_0)  \\ &&+\tanh(2 q_l(\alpha^- v^{**}- \alpha^+ v^*-0.5 \eta)-y_0))] \\ && +q_l\lambda_2 (\hat{K}_l^+(k) +\hat{K}_l^-(k)) \{ v^*[1-\tanh^2(2 q_l( \alpha^- v^{**}-\alpha^+ v^*-0.5 \eta)-y_0)] \\ && + v^{**}[1-\tanh^2(-2 q_l(\alpha^- v^{**}-\alpha^+ v^*-0.5 \eta)-y_0)]\}, \\
D(k)&=& 4 \beta_+ \beta_- k^2 +4ik\lambda_1(\beta_+-\beta_-)+ 2ik \lambda_2 \{ \beta_+(1+\tanh(-2q_l(\alpha^- v^{**}-\alpha^+ v^*-0.5\eta) -y_0)) \\ && -\beta_-(1+\tanh(2q_l(\alpha^- v^{**}-\alpha^+ v^*-0.5\eta) -y_0)) \\ &&
+v^*(1-\tanh^2(2q_l(\alpha^- v^{**}-\alpha^+ v^*-0.5\eta)-y_0))[\beta_+(q_a \hat{K}_a^+ + q_l(\hat{K}_l^+ + \hat{K}_l^-)) \\ &&+ \beta_-(q_a \hat{K}_a^+ - q_l(\hat{K}_l^+ + \hat{K}_l^-))] \\&&  + v^{**}(1-\tanh^2(-2q_l(\alpha^- v^{**}-\alpha^+ v^*-0.5\eta) -y_0))[\beta_+(-q_a \hat{K}_a^- - q_l(\hat{K}_l^+ + \hat{K}_l^-)) \\&& + \beta_-(-q_a \hat{K}_a^- + q_l(\hat{K}_l^+ + \hat{K}_l^-))] \}.
\end{eqnarray*}

In the above, $\hat{K}^\pm_j(k)$, for $j=a, l$ denote the Fourier transform of the kernel $K_j(s)$ defined in Eq. \eqref{kernels}, i.e. 
\begin{equation*}
\hat{K}^\pm_j(k)=\int_{-\infty}^{+\infty} K_j(s) e^{\pm iks} \textup{d}s = \exp\left( \pm is_j k -\frac{k^2 m_l^2}{2} \right), \quad j=a,l.     
\end{equation*}

\begin{figure}[htb]
    \centering
\includegraphics[width=.7\textwidth]{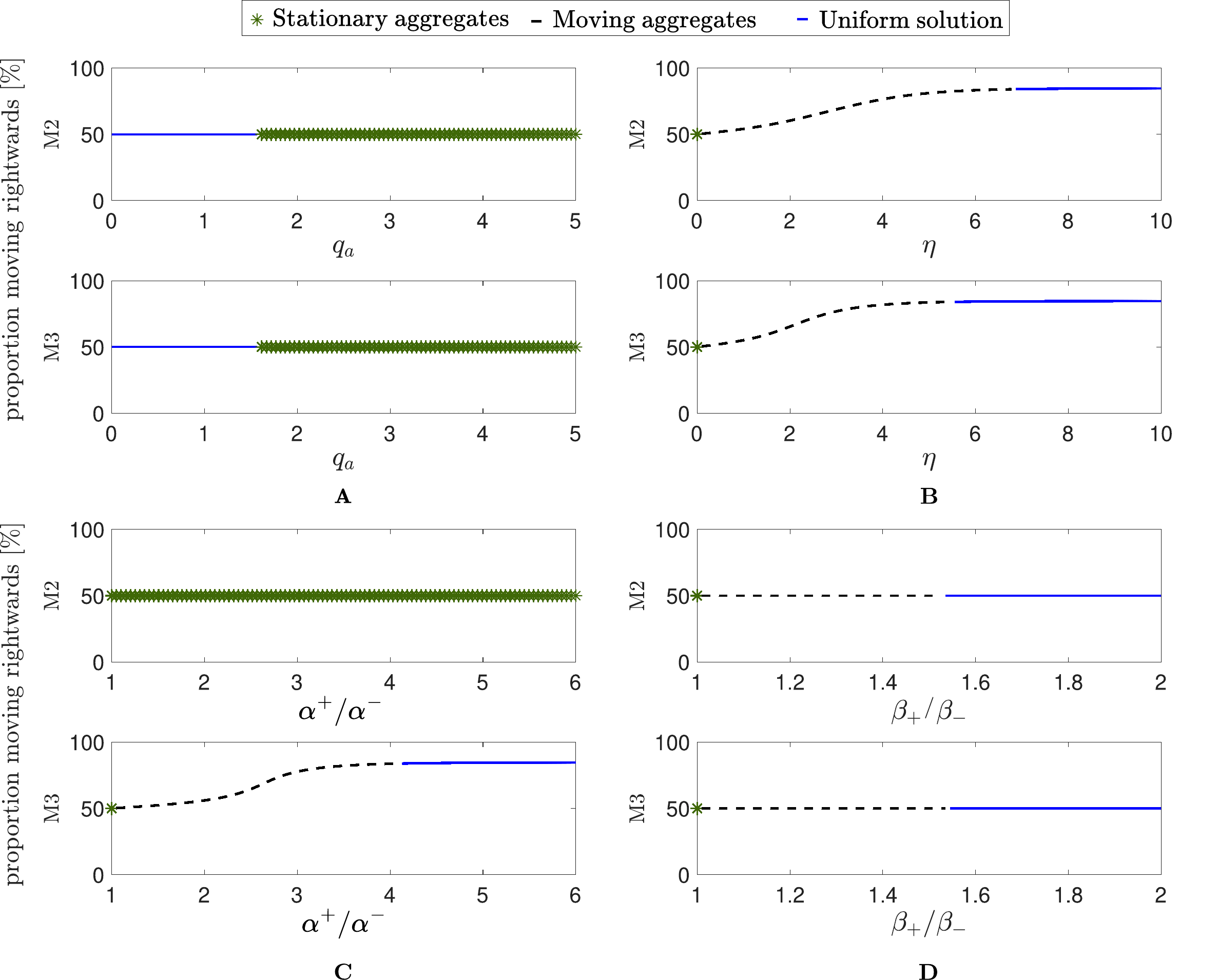}
    \caption{M2 and M3 leader-only sub-models, Proportion of right-moving populations at steady state. Effect of attraction strength $q_a$, $\eta$ (Obias strength), $\alpha^+/\alpha^-$ (Cbias strength) and $\beta_+/\beta_-$ (Sbias strength) on position and linear stability properties of steady states. Other parameter values: $q_l=0.5$, $q_a=2$ in (B, C, D), $\lambda_1=0.2$, and $\lambda_2=0.9$.}
    \label{LSA_M2_M3}
\end{figure}

Figure \ref{LSA_M2_M3} outlines the key insights that we can get from linear stability analysis of the leader-only submodels. A certain amount of attraction is crucial to aggregate a dispersed population (see Figure \ref{LSA_M2_M3}A). Assuming sufficient attraction, the introduction of Obias generates consensus of the migration towards the target for both M2 and M3 submodels (see Figure \ref{LSA_M2_M3}B). 
The M2 leader-submodel does not depend on the Cbias parameters (i.e. $\alpha^+/\alpha^-$). Consistently, Cbias strategy has no effect in the M2 leader-only submodel. Indeed, in the \emph{observant} leader model Cbias only affect the alignment of the followers, which are not taken into account here.
M3 leader-submodel instead generates a shift of the group orientation towards the target (see Figure \ref{LSA_M2_M3}C). Finally, variation in the ratio $\beta_+/\beta_-$ does not alter the steady state but does affect its stability (see Figure \ref{LSA_M2_M3}D).
Excessively high values of the bias strength $\eta$, $\alpha^+/\alpha^-$, $\beta_+/\beta_-$ result in the population dispersing towards the uniform solution.  


\section{Fixed parameters}\label{fixedpar}
Table \ref{table_fixed} summarizes the parameters set at a fixed reference value, in accordance to the previous studies in \cite{eftimie2007modeling,bernardi2021leadership}. 

\begin{table}[h] 
\caption{Table of parameters kept fixed within this study.}
\label{table_fixed}       
\begin{tabular}{llll}
\multicolumn{1}{c}{\textbf{Grouping}} & \multicolumn{1}{c}{\textbf{Parameter}} & \multicolumn{1}{c}{\textbf{Description}} & \multicolumn{1}{c}{\textbf{Value [Unit]}}\\
    \hline
    \\
Speed: &$\gamma$          & follower speed &0.1 [L/T]\\   
\\
              \hline
\\
Interaction kernels: 
                     &$s_l$ &alignment range  &0.5 [L]\\
                   &$s_a$ &attraction range &1[L]\\
                   &$m_l$ &width of alignment kernel &0.5/8 [L]\\
                   &$m_a$ &width of attraction kernel &1/8 [L]\\
\\
              \hline
\\
Turning rates: &$y_0$     &shift of the turning function &2\\
\noalign{\smallskip}\hline
\end{tabular}
\end{table}

\section{Extra Figures}
\label{appendixExtraFigures}

\begin{figure}[htb]
    \centering
\includegraphics[width=\textwidth]{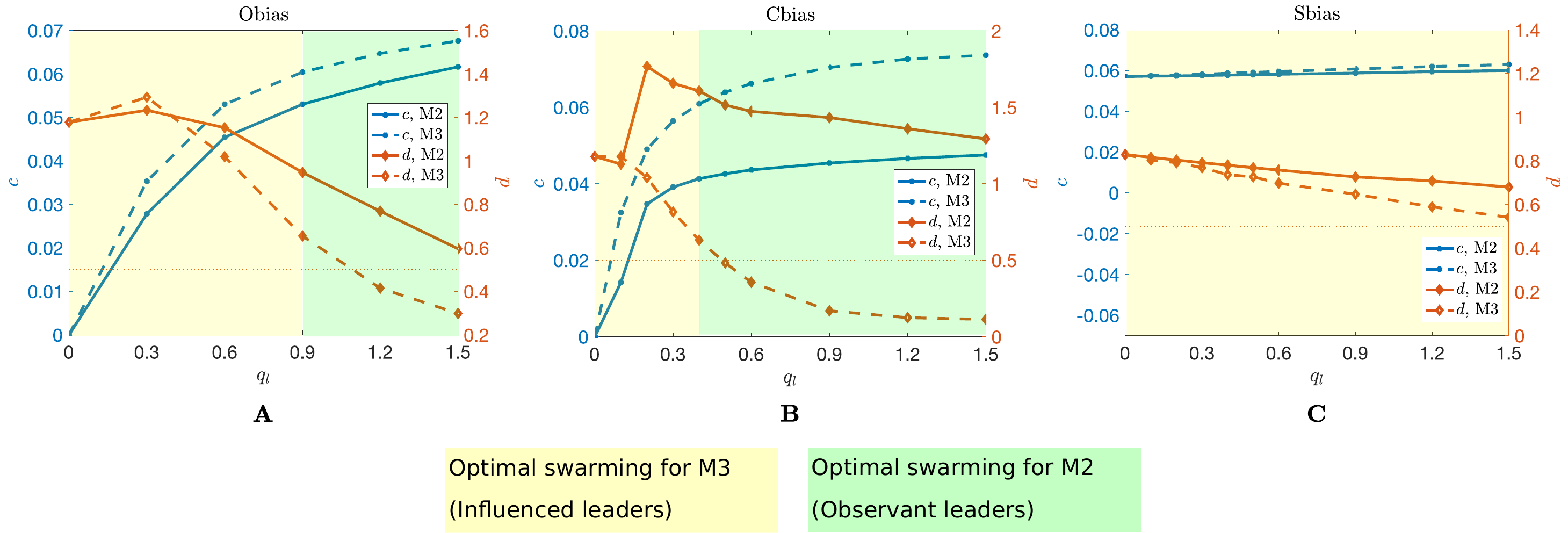}
    \caption{M2 and M3, Effect of Obias, Sbias and Cbias leader strategies on the swarm dynamics as $q_l$ increases under high attraction regime
 and clustered initial configuration. Numerical simulations are obtained for (A) $\eta=10$, $\alpha^\pm=1$, $\beta_\pm=0.1$, (B) $\beta_+/\beta_-=0.5/0.1$, $\eta=1$, $\alpha^\pm=1$, (C) $\alpha^+/\alpha^-=5/1$, $\beta_\pm=0.1$, $\eta=1$.
 Swarm dynamics is evaluated in terms of the speed (in blue) and the cohesion index (in orange) of the follower population. In yellow and green, we highlight the $q_l$ values for which optimal swarming is obtained for M3 and M2, respectively. Other parameter values are set as $q_a=2$, $\lambda_1=0.2$, $\lambda_2=0.9$, $M_u=12.61$, $M_v=12.61$, and $x_0=5$.}
    \label{results_M2_M3_qa_2}
\end{figure}

\end{appendices}

\bibliography{sn-article}


\begin{thebibliography}{40}
\ifx \bisbn   \undefined \def \bisbn  #1{ISBN #1}\fi
\ifx \binits  \undefined \def \binits#1{#1}\fi
\ifx \bauthor  \undefined \def \bauthor#1{#1}\fi
\ifx \batitle  \undefined \def \batitle#1{#1}\fi
\ifx \bjtitle  \undefined \def \bjtitle#1{#1}\fi
\ifx \bvolume  \undefined \def \bvolume#1{\textbf{#1}}\fi
\ifx \byear  \undefined \def \byear#1{#1}\fi
\ifx \bissue  \undefined \def \bissue#1{#1}\fi
\ifx \bfpage  \undefined \def \bfpage#1{#1}\fi
\ifx \blpage  \undefined \def \blpage #1{#1}\fi
\ifx \burl  \undefined \def \burl#1{\textsf{#1}}\fi
\ifx \doiurl  \undefined \def \doiurl#1{\url{https://doi.org/#1}}\fi
\ifx \betal  \undefined \def \betal{\textit{et al.}}\fi
\ifx \binstitute  \undefined \def \binstitute#1{#1}\fi
\ifx \binstitutionaled  \undefined \def \binstitutionaled#1{#1}\fi
\ifx \bctitle  \undefined \def \bctitle#1{#1}\fi
\ifx \beditor  \undefined \def \beditor#1{#1}\fi
\ifx \bpublisher  \undefined \def \bpublisher#1{#1}\fi
\ifx \bbtitle  \undefined \def \bbtitle#1{#1}\fi
\ifx \bedition  \undefined \def \bedition#1{#1}\fi
\ifx \bseriesno  \undefined \def \bseriesno#1{#1}\fi
\ifx \blocation  \undefined \def \blocation#1{#1}\fi
\ifx \bsertitle  \undefined \def \bsertitle#1{#1}\fi
\ifx \bsnm \undefined \def \bsnm#1{#1}\fi
\ifx \bsuffix \undefined \def \bsuffix#1{#1}\fi
\ifx \bparticle \undefined \def \bparticle#1{#1}\fi
\ifx \barticle \undefined \def \barticle#1{#1}\fi
\bibcommenthead
\ifx \bconfdate \undefined \def \bconfdate #1{#1}\fi
\ifx \botherref \undefined \def \botherref #1{#1}\fi
\ifx \url \undefined \def \url#1{\textsf{#1}}\fi
\ifx \bchapter \undefined \def \bchapter#1{#1}\fi
\ifx \bbook \undefined \def \bbook#1{#1}\fi
\ifx \bcomment \undefined \def \bcomment#1{#1}\fi
\ifx \oauthor \undefined \def \oauthor#1{#1}\fi
\ifx \citeauthoryear \undefined \def \citeauthoryear#1{#1}\fi
\ifx \endbibitem  \undefined \def \endbibitem {}\fi
\ifx \bconflocation  \undefined \def \bconflocation#1{#1}\fi
\ifx \arxivurl  \undefined \def \arxivurl#1{\textsf{#1}}\fi
\csname PreBibitemsHook\endcsname

\bibitem[\protect\citeauthoryear{Lewis}{1994}]{lewis1994we}
\begin{bbook}
\bauthor{\bsnm{Lewis}, \binits{D.}}:
\bbtitle{We, the Navigators: The Ancient Art of Landfinding in the Pacific},
\bedition{2}nd edn.
\bpublisher{University of Hawaii Press},
\blocation{Honolulu}
(\byear{1994})
\end{bbook}
\endbibitem

\bibitem[\protect\citeauthoryear{Berdahl et~al.}{2018}]{berdahl2018collective}
\begin{barticle}
\bauthor{\bsnm{Berdahl}, \binits{A.M.}},
\bauthor{\bsnm{Kao}, \binits{A.B.}},
\bauthor{\bsnm{Flack}, \binits{A.}},
\bauthor{\bsnm{Westley}, \binits{P.A.}},
\bauthor{\bsnm{Codling}, \binits{E.A.}},
\bauthor{\bsnm{Couzin}, \binits{I.D.}},
\bauthor{\bsnm{Dell}, \binits{A.I.}},
\bauthor{\bsnm{Biro}, \binits{D.}}:
\batitle{Collective animal navigation and migratory culture: from theoretical
  models to empirical evidence}.
\bjtitle{Philosophical Transactions of the Royal Society B: Biological
  Sciences}
\bvolume{373}(\bissue{1746}),
\bfpage{20170009}
(\byear{2018})
\end{barticle}
\endbibitem

\bibitem[\protect\citeauthoryear{Brent et~al.}{2015}]{brent2015ecological}
\begin{barticle}
\bauthor{\bsnm{Brent}, \binits{L.J.}},
\bauthor{\bsnm{Franks}, \binits{D.W.}},
\bauthor{\bsnm{Foster}, \binits{E.A.}},
\bauthor{\bsnm{Balcomb}, \binits{K.C.}},
\bauthor{\bsnm{Cant}, \binits{M.A.}},
\bauthor{\bsnm{Croft}, \binits{D.P.}}:
\batitle{Ecological knowledge, leadership, and the evolution of menopause in
  killer whales}.
\bjtitle{Current Biology}
\bvolume{25}(\bissue{6}),
\bfpage{746}--\blpage{750}
(\byear{2015})
\end{barticle}
\endbibitem

\bibitem[\protect\citeauthoryear{Strandburg-Peshkin
  et~al.}{2018}]{strandburg2018inferring}
\begin{barticle}
\bauthor{\bsnm{Strandburg-Peshkin}, \binits{A.}},
\bauthor{\bsnm{Papageorgiou}, \binits{D.}},
\bauthor{\bsnm{Crofoot}, \binits{M.C.}},
\bauthor{\bsnm{Farine}, \binits{D.R.}}:
\batitle{Inferring influence and leadership in moving animal groups}.
\bjtitle{Philosophical Transactions of the Royal Society B: Biological
  Sciences}
\bvolume{373}(\bissue{1746}),
\bfpage{20170006}
(\byear{2018})
\end{barticle}
\endbibitem

\bibitem[\protect\citeauthoryear{Vishwakarma
  et~al.}{2018}]{vishwakarma2018mechanical}
\begin{barticle}
\bauthor{\bsnm{Vishwakarma}, \binits{M.}},
\bauthor{\bsnm{Di~Russo}, \binits{J.}},
\bauthor{\bsnm{Probst}, \binits{D.}},
\bauthor{\bsnm{Schwarz}, \binits{U.S.}},
\bauthor{\bsnm{Das}, \binits{T.}},
\bauthor{\bsnm{Spatz}, \binits{J.P.}}:
\batitle{Mechanical interactions among followers determine the emergence of
  leaders in migrating epithelial cell collectives}.
\bjtitle{Nature Communications}
\bvolume{9}(\bissue{1}),
\bfpage{1}--\blpage{12}
(\byear{2018})
\end{barticle}
\endbibitem

\bibitem[\protect\citeauthoryear{Cheung et~al.}{2013}]{cheung2013collective}
\begin{barticle}
\bauthor{\bsnm{Cheung}, \binits{K.J.}},
\bauthor{\bsnm{Gabrielson}, \binits{E.}},
\bauthor{\bsnm{Werb}, \binits{Z.}},
\bauthor{\bsnm{Ewald}, \binits{A.J.}}:
\batitle{Collective invasion in breast cancer requires a conserved basal
  epithelial program}.
\bjtitle{Cell}
\bvolume{155}(\bissue{7}),
\bfpage{1639}--\blpage{1651}
(\byear{2013})
\end{barticle}
\endbibitem

\bibitem[\protect\citeauthoryear{Vilchez~Mercedes
  et~al.}{2021}]{vilchez2021decoding}
\begin{barticle}
\bauthor{\bsnm{Vilchez~Mercedes}, \binits{S.A.}},
\bauthor{\bsnm{Bocci}, \binits{F.}},
\bauthor{\bsnm{Levine}, \binits{H.}},
\bauthor{\bsnm{Onuchic}, \binits{J.N.}},
\bauthor{\bsnm{Jolly}, \binits{M.K.}},
\bauthor{\bsnm{Wong}, \binits{P.K.}}:
\batitle{Decoding leader cells in collective cancer invasion}.
\bjtitle{Nature Reviews Cancer}
\bvolume{21}(\bissue{9}),
\bfpage{592}--\blpage{604}
(\byear{2021})
\end{barticle}
\endbibitem

\bibitem[\protect\citeauthoryear{Payne}{2003}]{payne2003sources}
\begin{bchapter}
\bauthor{\bsnm{Payne}, \binits{K.}}:
\bctitle{Sources of social complexity in the three elephant species}.
In: \beditor{\bsnm{Waal}, \binits{F.B.M.}},
\beditor{\bsnm{Tyack}, \binits{P.L.}} (eds.)
\bbtitle{Animal Social Complexity},
pp. \bfpage{57}--\blpage{86}.
\bpublisher{Harvard University Press},
\blocation{Cambridge, MA and London, England}
(\byear{2003})
\end{bchapter}
\endbibitem

\bibitem[\protect\citeauthoryear{Averly et~al.}{2022}]{averly2022disentangling}
\begin{barticle}
\bauthor{\bsnm{Averly}, \binits{B.}},
\bauthor{\bsnm{Sridhar}, \binits{V.H.}},
\bauthor{\bsnm{Demartsev}, \binits{V.}},
\bauthor{\bsnm{Gall}, \binits{G.}},
\bauthor{\bsnm{Manser}, \binits{M.}},
\bauthor{\bsnm{Strandburg-Peshkin}, \binits{A.}}:
\batitle{Disentangling influence over group speed and direction reveals
  multiple patterns of influence in moving meerkat groups}.
\bjtitle{Scientific Reports}
\bvolume{12}(\bissue{1}),
\bfpage{13844}
(\byear{2022})
\end{barticle}
\endbibitem

\bibitem[\protect\citeauthoryear{Sasaki et~al.}{2018}]{sasaki2018personality}
\begin{barticle}
\bauthor{\bsnm{Sasaki}, \binits{T.}},
\bauthor{\bsnm{Mann}, \binits{R.P.}},
\bauthor{\bsnm{Warren}, \binits{K.N.}},
\bauthor{\bsnm{Herbert}, \binits{T.}},
\bauthor{\bsnm{Wilson}, \binits{T.}},
\bauthor{\bsnm{Biro}, \binits{D.}}:
\batitle{Personality and the collective: bold homing pigeons occupy higher
  leadership ranks in flocks}.
\bjtitle{Philosophical Transactions of the Royal Society B: Biological
  Sciences}
\bvolume{373}(\bissue{1746}),
\bfpage{20170038}
(\byear{2018})
\end{barticle}
\endbibitem

\bibitem[\protect\citeauthoryear{Webster et~al.}{2017}]{webster2017fish}
\begin{barticle}
\bauthor{\bsnm{Webster}, \binits{M.M.}},
\bauthor{\bsnm{Whalen}, \binits{A.}},
\bauthor{\bsnm{Laland}, \binits{K.N.}}:
\batitle{Fish pool their experience to solve problems collectively}.
\bjtitle{Nature Ecology \& Evolution}
\bvolume{1}(\bissue{5}),
\bfpage{1}--\blpage{5}
(\byear{2017})
\end{barticle}
\endbibitem

\bibitem[\protect\citeauthoryear{Aoki}{1982}]{aoki1982}
\begin{botherref}
\oauthor{\bsnm{Aoki}, \binits{I.}}:
A simulation study on the schooling mechanism in fish.
Bulletin of the Japanese Society of Scientific Fisheries (Japan)
\textbf{48}(8)
(1982)
\end{botherref}
\endbibitem

\bibitem[\protect\citeauthoryear{Reynolds}{1987}]{reynolds1987flocks}
\begin{bchapter}
\bauthor{\bsnm{Reynolds}, \binits{C.W.}}:
\bctitle{Flocks, herds and schools: A distributed behavioral model}.
In: \bbtitle{Proceedings of the 14th Annual Conference on Computer Graphics and
  Interactive Techniques},
pp. \bfpage{25}--\blpage{34}
(\byear{1987})
\end{bchapter}
\endbibitem

\bibitem[\protect\citeauthoryear{Vicsek et~al.}{1995}]{vicsek1995novel}
\begin{barticle}
\bauthor{\bsnm{Vicsek}, \binits{T.}},
\bauthor{\bsnm{Czir{\'o}k}, \binits{A.}},
\bauthor{\bsnm{Ben-Jacob}, \binits{E.}},
\bauthor{\bsnm{Cohen}, \binits{I.}},
\bauthor{\bsnm{Shochet}, \binits{O.}}:
\batitle{Novel type of phase transition in a system of self-driven particles}.
\bjtitle{Physical Review Letters}
\bvolume{75}(\bissue{6}),
\bfpage{1226}
(\byear{1995})
\end{barticle}
\endbibitem

\bibitem[\protect\citeauthoryear{Couzin et~al.}{2002}]{couzin2002collective}
\begin{barticle}
\bauthor{\bsnm{Couzin}, \binits{I.D.}},
\bauthor{\bsnm{Krause}, \binits{J.}},
\bauthor{\bsnm{James}, \binits{R.}},
\bauthor{\bsnm{Ruxton}, \binits{G.D.}},
\bauthor{\bsnm{Franks}, \binits{N.R.}}:
\batitle{Collective memory and spatial sorting in animal groups}.
\bjtitle{Journal of Theoretical Biology}
\bvolume{218}(\bissue{1}),
\bfpage{1}--\blpage{11}
(\byear{2002})
\end{barticle}
\endbibitem

\bibitem[\protect\citeauthoryear{Eftimie}{2018}]{eftimie2018hyperbolic}
\begin{botherref}
\oauthor{\bsnm{Eftimie}, \binits{R.}}:
Hyperbolic and kinetic models for self-organised biological aggregations.
A modelling and pattern formation approach. Cham, Switzerland: Springer
(2018)
\end{botherref}
\endbibitem

\bibitem[\protect\citeauthoryear{Painter et~al.}{2024}]{painter2024biological}
\begin{barticle}
\bauthor{\bsnm{Painter}, \binits{K.J.}},
\bauthor{\bsnm{Hillen}, \binits{T.}},
\bauthor{\bsnm{Potts}, \binits{J.R.}}:
\batitle{Biological modeling with nonlocal advection--diffusion equations}.
\bjtitle{Mathematical Models and Methods in Applied Sciences}
\bvolume{34}(\bissue{01}),
\bfpage{57}--\blpage{107}
(\byear{2024})
\end{barticle}
\endbibitem

\bibitem[\protect\citeauthoryear{Carrillo et~al.}{2010}]{carrillo2010particle}
\begin{bbook}
\bauthor{\bsnm{Carrillo}, \binits{J.A.}},
\bauthor{\bsnm{Fornasier}, \binits{M.}},
\bauthor{\bsnm{Toscani}, \binits{G.}},
\bauthor{\bsnm{Vecil}, \binits{F.}}:
In: \beditor{\bsnm{Naldi}, \binits{G.}},
\beditor{\bsnm{Pareschi}, \binits{L.}},
\beditor{\bsnm{Toscani}, \binits{G.}} (eds.)
\bbtitle{Particle, kinetic, and hydrodynamic models of swarming},
pp. \bfpage{297}--\blpage{336}.
\bpublisher{Birkh{\"a}user Boston},
\blocation{Boston}
(\byear{2010})
\end{bbook}
\endbibitem

\bibitem[\protect\citeauthoryear{Codling and Bode}{2016}]{codling2016balancing}
\begin{barticle}
\bauthor{\bsnm{Codling}, \binits{E.A.}},
\bauthor{\bsnm{Bode}, \binits{N.W.}}:
\batitle{Balancing direct and indirect sources of navigational information in a
  leaderless model of collective animal movement}.
\bjtitle{Journal of Theoretical Biology}
\bvolume{394},
\bfpage{32}--\blpage{42}
(\byear{2016})
\end{barticle}
\endbibitem

\bibitem[\protect\citeauthoryear{Saltz et~al.}{2023}]{saltz2023identifying}
\begin{barticle}
\bauthor{\bsnm{Saltz}, \binits{J.}},
\bauthor{\bsnm{Palmer}, \binits{M.}},
\bauthor{\bsnm{Beaudrot}, \binits{L.}}:
\batitle{Identifying the social context of single-and mixed-species group
  formation in large {A}frican herbivores}.
\bjtitle{Philosophical Transactions of the Royal Society B}
\bvolume{378}(\bissue{1878}),
\bfpage{20220105}
(\byear{2023})
\end{barticle}
\endbibitem

\bibitem[\protect\citeauthoryear{Giese et~al.}{1996}]{giese1996dichotomy}
\begin{barticle}
\bauthor{\bsnm{Giese}, \binits{A.}},
\bauthor{\bsnm{Loo}, \binits{M.A.}},
\bauthor{\bsnm{Tran}, \binits{N.}},
\bauthor{\bsnm{Haskett}, \binits{D.}},
\bauthor{\bsnm{Coons}, \binits{S.W.}},
\bauthor{\bsnm{Berens}, \binits{M.E.}}:
\batitle{Dichotomy of astrocytoma migration and proliferation}.
\bjtitle{International Journal of Cancer}
\bvolume{67}(\bissue{2}),
\bfpage{275}--\blpage{282}
(\byear{1996})
\end{barticle}
\endbibitem

\bibitem[\protect\citeauthoryear{Guttal and Couzin}{2010}]{guttal2010social}
\begin{barticle}
\bauthor{\bsnm{Guttal}, \binits{V.}},
\bauthor{\bsnm{Couzin}, \binits{I.D.}}:
\batitle{Social interactions, information use, and the evolution of collective
  migration}.
\bjtitle{Proceedings of the National Academy of Sciences}
\bvolume{107}(\bissue{37}),
\bfpage{16172}--\blpage{16177}
(\byear{2010})
\end{barticle}
\endbibitem

\bibitem[\protect\citeauthoryear{Couzin et~al.}{2011}]{couzin2011uninformed}
\begin{barticle}
\bauthor{\bsnm{Couzin}, \binits{I.D.}},
\bauthor{\bsnm{Ioannou}, \binits{C.C.}},
\bauthor{\bsnm{Demirel}, \binits{G.}},
\bauthor{\bsnm{Gross}, \binits{T.}},
\bauthor{\bsnm{Torney}, \binits{C.J.}},
\bauthor{\bsnm{Hartnett}, \binits{A.}},
\bauthor{\bsnm{Conradt}, \binits{L.}},
\bauthor{\bsnm{Levin}, \binits{S.A.}},
\bauthor{\bsnm{Leonard}, \binits{N.E.}}:
\batitle{Uninformed individuals promote democratic consensus in animal groups}.
\bjtitle{Science}
\bvolume{334}(\bissue{6062}),
\bfpage{1578}--\blpage{1580}
(\byear{2011})
\end{barticle}
\endbibitem

\bibitem[\protect\citeauthoryear{Bernardi
  et~al.}{2021}]{bernardi2021leadership}
\begin{barticle}
\bauthor{\bsnm{Bernardi}, \binits{S.}},
\bauthor{\bsnm{Eftimie}, \binits{R.}},
\bauthor{\bsnm{Painter}, \binits{K.J.}}:
\batitle{Leadership through influence: what mechanisms allow leaders to steer a
  swarm?}
\bjtitle{Bulletin of Mathematical Biology}
\bvolume{83}(\bissue{6}),
\bfpage{1}--\blpage{33}
(\byear{2021})
\end{barticle}
\endbibitem

\bibitem[\protect\citeauthoryear{Eftimie et~al.}{2007}]{eftimie2007modeling}
\begin{barticle}
\bauthor{\bsnm{Eftimie}, \binits{R.}},
\bauthor{\bsnm{De~Vries}, \binits{G.}},
\bauthor{\bsnm{Lewis}, \binits{M.A.}},
\bauthor{\bsnm{Lutscher}, \binits{F.}}:
\batitle{Modeling group formation and activity patterns in self-organizing
  collectives of individuals}.
\bjtitle{Bulletin of Mathematical Biology}
\bvolume{69}(\bissue{5}),
\bfpage{1537}--\blpage{1565}
(\byear{2007})
\end{barticle}
\endbibitem

\bibitem[\protect\citeauthoryear{Qin et~al.}{2021}]{qin2021roles}
\begin{barticle}
\bauthor{\bsnm{Qin}, \binits{L.}},
\bauthor{\bsnm{Yang}, \binits{D.}},
\bauthor{\bsnm{Yi}, \binits{W.}},
\bauthor{\bsnm{Cao}, \binits{H.}},
\bauthor{\bsnm{Xiao}, \binits{G.}}:
\batitle{Roles of leader and follower cells in collective cell migration}.
\bjtitle{Molecular Biology of the Cell}
\bvolume{32}(\bissue{14}),
\bfpage{1267}--\blpage{1272}
(\byear{2021})
\end{barticle}
\endbibitem

\bibitem[\protect\citeauthoryear{Jolles et~al.}{2017}]{jolles2017consistent}
\begin{barticle}
\bauthor{\bsnm{Jolles}, \binits{J.W.}},
\bauthor{\bsnm{Boogert}, \binits{N.J.}},
\bauthor{\bsnm{Sridhar}, \binits{V.H.}},
\bauthor{\bsnm{Couzin}, \binits{I.D.}},
\bauthor{\bsnm{Manica}, \binits{A.}}:
\batitle{Consistent individual differences drive collective behavior and group
  functioning of schooling fish}.
\bjtitle{Current Biology}
\bvolume{27}(\bissue{18}),
\bfpage{2862}--\blpage{2868}
(\byear{2017})
\end{barticle}
\endbibitem

\bibitem[\protect\citeauthoryear{Pettit et~al.}{2015}]{pettit2015speed}
\begin{barticle}
\bauthor{\bsnm{Pettit}, \binits{B.}},
\bauthor{\bsnm{Akos}, \binits{Z.}},
\bauthor{\bsnm{Vicsek}, \binits{T.}},
\bauthor{\bsnm{Biro}, \binits{D.}}:
\batitle{Speed determines leadership and leadership determines learning during
  pigeon flocking}.
\bjtitle{Current Biology}
\bvolume{25}(\bissue{23}),
\bfpage{3132}--\blpage{3137}
(\byear{2015})
\end{barticle}
\endbibitem

\bibitem[\protect\citeauthoryear{Ioannou et~al.}{2015}]{ioannou2015potential}
\begin{barticle}
\bauthor{\bsnm{Ioannou}, \binits{C.C.}},
\bauthor{\bsnm{Singh}, \binits{M.}},
\bauthor{\bsnm{Couzin}, \binits{I.D.}}:
\batitle{Potential leaders trade off goal-oriented and socially oriented
  behavior in mobile animal groups}.
\bjtitle{The American Naturalist}
\bvolume{186}(\bissue{2}),
\bfpage{284}--\blpage{293}
(\byear{2015})
\end{barticle}
\endbibitem

\bibitem[\protect\citeauthoryear{Miller et~al.}{2013}]{miller2013both}
\begin{barticle}
\bauthor{\bsnm{Miller}, \binits{N.}},
\bauthor{\bsnm{Garnier}, \binits{S.}},
\bauthor{\bsnm{Hartnett}, \binits{A.T.}},
\bauthor{\bsnm{Couzin}, \binits{I.D.}}:
\batitle{Both information and social cohesion determine collective decisions in
  animal groups}.
\bjtitle{Proceedings of the National Academy of Sciences}
\bvolume{110}(\bissue{13}),
\bfpage{5263}--\blpage{5268}
(\byear{2013})
\end{barticle}
\endbibitem

\bibitem[\protect\citeauthoryear{Schultz et~al.}{2008}]{schultz2008mechanism}
\begin{barticle}
\bauthor{\bsnm{Schultz}, \binits{K.M.}},
\bauthor{\bsnm{Passino}, \binits{K.M.}},
\bauthor{\bsnm{Seeley}, \binits{T.D.}}:
\batitle{The mechanism of flight guidance in honeybee swarms: subtle guides or
  streaker bees?}
\bjtitle{Journal of Experimental Biology}
\bvolume{211}(\bissue{20}),
\bfpage{3287}--\blpage{3295}
(\byear{2008})
\end{barticle}
\endbibitem

\bibitem[\protect\citeauthoryear{Herbert-Read}{2015}]{herbert2015collective}
\begin{barticle}
\bauthor{\bsnm{Herbert-Read}, \binits{J.}}:
\batitle{Collective behaviour: leadership and learning in flocks}.
\bjtitle{Current Biology}
\bvolume{25}(\bissue{23}),
\bfpage{1127}--\blpage{1129}
(\byear{2015})
\end{barticle}
\endbibitem

\bibitem[\protect\citeauthoryear{Jolles et~al.}{2020}]{jolles2020role}
\begin{barticle}
\bauthor{\bsnm{Jolles}, \binits{J.W.}},
\bauthor{\bsnm{King}, \binits{A.J.}},
\bauthor{\bsnm{Killen}, \binits{S.S.}}:
\batitle{The role of individual heterogeneity in collective animal behaviour}.
\bjtitle{Trends in Ecology \& Evolution}
\bvolume{35}(\bissue{3}),
\bfpage{278}--\blpage{291}
(\byear{2020})
\end{barticle}
\endbibitem

\bibitem[\protect\citeauthoryear{McLennan
  et~al.}{2012}]{mclennan2012multiscale}
\begin{barticle}
\bauthor{\bsnm{McLennan}, \binits{R.}},
\bauthor{\bsnm{Dyson}, \binits{L.}},
\bauthor{\bsnm{Prather}, \binits{K.W.}},
\bauthor{\bsnm{Morrison}, \binits{J.A.}},
\bauthor{\bsnm{Baker}, \binits{R.E.}},
\bauthor{\bsnm{Maini}, \binits{P.K.}},
\bauthor{\bsnm{Kulesa}, \binits{P.M.}}:
\batitle{Multiscale mechanisms of cell migration during development: theory and
  experiment}.
\bjtitle{Development}
\bvolume{139}(\bissue{16}),
\bfpage{2935}--\blpage{2944}
(\byear{2012})
\end{barticle}
\endbibitem

\bibitem[\protect\citeauthoryear{McLennan et~al.}{2015}]{mclennan2015vegf}
\begin{barticle}
\bauthor{\bsnm{McLennan}, \binits{R.}},
\bauthor{\bsnm{Schumacher}, \binits{L.J.}},
\bauthor{\bsnm{Morrison}, \binits{J.A.}},
\bauthor{\bsnm{Teddy}, \binits{J.M.}},
\bauthor{\bsnm{Ridenour}, \binits{D.A.}},
\bauthor{\bsnm{Box}, \binits{A.C.}},
\bauthor{\bsnm{Semerad}, \binits{C.L.}},
\bauthor{\bsnm{Li}, \binits{H.}},
\bauthor{\bsnm{McDowell}, \binits{W.}},
\bauthor{\bsnm{Kay}, \binits{D.}},
\bauthor{\bsnm{Maini}, \binits{P.K.}},
\bauthor{\bsnm{Baker}, \binits{R.E.}},
\bauthor{\bsnm{Kulesa}, \binits{P.M.}}:
\batitle{{VEGF} signals induce trailblazer cell identity that drives neural
  crest migration}.
\bjtitle{Developmental Biology}
\bvolume{407}(\bissue{1}),
\bfpage{12}--\blpage{25}
(\byear{2015})
\end{barticle}
\endbibitem

\bibitem[\protect\citeauthoryear{Albi and Ferrarese}{2024}]{albi2024kinetic}
\begin{barticle}
\bauthor{\bsnm{Albi}, \binits{G.}},
\bauthor{\bsnm{Ferrarese}, \binits{F.}}:
\batitle{Kinetic description of swarming dynamics with topological interaction
  and transient leaders}.
\bjtitle{Multiscale Modeling \& Simulation}
\bvolume{22}(\bissue{3}),
\bfpage{1169}--\blpage{1195}
(\byear{2024})
\end{barticle}
\endbibitem

\bibitem[\protect\citeauthoryear{Cristiani et~al.}{2025}]{cristiani2024kinetic}
\begin{barticle}
\bauthor{\bsnm{Cristiani}, \binits{E.}},
\bauthor{\bsnm{Loy}, \binits{N.}},
\bauthor{\bsnm{Menci}, \binits{M.}},
\bauthor{\bsnm{Tosin}, \binits{A.}}:
\batitle{Kinetic description and macroscopic limit of swarming dynamics with
  continuous leader–follower transitions}.
\bjtitle{Mathematics and Computers in Simulation}
\bvolume{228},
\bfpage{362}--\blpage{385}
(\byear{2025})
\end{barticle}
\endbibitem

\bibitem[\protect\citeauthoryear{Painter et~al.}{2024}]{painter2024variations}
\begin{botherref}
\oauthor{\bsnm{Painter}, \binits{K.J.}},
\oauthor{\bsnm{Giunta}, \binits{V.}},
\oauthor{\bsnm{Potts}, \binits{J.R.}},
\oauthor{\bsnm{Bernardi}, \binits{S.}}:
Variations in nonlocal interaction range lead to emergent chase-and-run in
  heterogeneous populations.
bioRxiv,
2024--06
(2024)
\end{botherref}
\endbibitem

\bibitem[\protect\citeauthoryear{Couzin et~al.}{2005}]{couzin2005effective}
\begin{barticle}
\bauthor{\bsnm{Couzin}, \binits{I.D.}},
\bauthor{\bsnm{Krause}, \binits{J.}},
\bauthor{\bsnm{Franks}, \binits{N.R.}},
\bauthor{\bsnm{Levin}, \binits{S.A.}}:
\batitle{Effective leadership and decision-making in animal groups on the
  move}.
\bjtitle{Nature}
\bvolume{433}(\bissue{7025}),
\bfpage{513}--\blpage{516}
(\byear{2005})
\end{barticle}
\endbibitem

\bibitem[\protect\citeauthoryear{Murray}{1993}]{murray2003mathematical}
\begin{botherref}
\oauthor{\bsnm{Murray}, \binits{J.D.}}:
Mathematical biology.
Springer-Verlag, Berlin
(1993)
\end{botherref}
\endbibitem

\end{thebibliography}

\end{document}